\documentclass[12pt]{spieman}  
\usepackage[]{aas_macros}  
\usepackage{hyperref}
\usepackage{amsmath,amsfonts,amssymb}
\usepackage{graphicx}
\usepackage{setspace}
\usepackage{tocloft}
\usepackage{wrapfig}
\usepackage[symbol]{footmisc}

\def\b1{\boldsymbol{1}}

\def\ba{\boldsymbol{a}}

\def\bc{\boldsymbol{c}}

\def\bD{\boldsymbol{D}}

\def\bE{\boldsymbol{E}}

\def\bg{{\bf g}}
\def\bg{\boldsymbol{g}}

\def\rm{\mathrm{m}}

\def\bs{\boldsymbol{s}}

\def\bU{\boldsymbol{U}}
\def\bu{\boldsymbol{u}}
\def\bv{\boldsymbol{v}}
\def\bV{\boldsymbol{V}}

\def\bW{\boldsymbol{W}}

\def\bx{\boldsymbol{x}}
\def\by{\boldsymbol{y}}

\def\bSigma{{\bf \Sigma}}

\newcommand{\bpac}{\big(\ba(\bc)\big)}

\title{Gratis Mitigation of Polarization Aberration Effects in Coronagraphic Dark Holes}

\author[$\dagger$]{Richard A. Frazin}
\affil[$\dagger$]{Dept. of Climate and Space Sciences, University of Michigan, Ann Arbor, MI 48109}

\cftpagenumbersoff{figure}
\cftpagenumbersoff{table} 
\begin{document} 
\maketitle


\begin{abstract}
	
Direct imaging of exoplanets requires stellar coronagraphs capable of suppressing starlight to contrast levels below $10^{-8}$.
Active wavefront control with deformable mirrors (DMs) is essential to create dark holes in the image plane.
However, polarization aberrations arising from beam reduction optics and the coronagraph itself produces multiple non-interfering intensity components that have correlated responses to the DM.
This article introduces the concept of gratis mitigation: when a control loop minimizes one intensity component, others can be reduced concomitantly due to the correlated DM responses.
Using end-to-end physical optics simulations of a Lyot coronagraph fed by a 4 m-class telescope with $f/5$ beam reduction, we demonstrate gratis mitigation and analyze its origin via a Jones matrix formalism.
Gratis mitigation has significant implications for coronagraph design, calibration, and possibly wavefront control.	

\end{abstract}


{\noindent \footnotesize\textbf{$\dagger$}Richard Frazin,  \linkable{rfrazin\emph{\_at\_}umich.edu} }

\section{Introduction}\label{sec: intro}

Direct imaging of exoplanets is among NASA's top priorities in future missions, such as the Habitable Worlds Observatory (HWO).
The principal challenge of direct imaging is the fact that exoplanets are located in close angular proximity to their vastly brighter host stars. 
Quantitatively, this means that the telescope optical system must be able to achieve a planet-to-star brightness ratio (i.e., \emph{contrast}), of less than $10^{-8}$, perhaps even $10^{-11}.$\cite{LUVOIR_model_JATIS22, Mennesson2024_HWOlab}
Current designs to meet this daunting requirement feature stellar coronagraphs, which suppress on-axis light while allowing slightly off-axis beams to pass through relatively unimpeded.\cite{Cavarroc_IdealCoronagraph06}
Even assuming perfect optical surfaces, diffraction alone would put the contrast levels orders of magnitude above the aforementioned requirements.
Real surfaces have aberrations at high and low spatial frequencies that further degrade the contrast.\cite{Krist_End2End_Roman_JATIS23}

Achieving high contrast requires active wavefront control strategies, notably the use of deformable mirrors (DMs).
In the context of space-based systems, the most prominent approach involving DMs is a family of techniques known collectively as \emph{electric field conjugation (EFC)}.
EFC procedures use one or two deformable mirrors (DMs) to modulate the intensity measured in the image plane through alternating sensing and control steps.
EFC procedures aim to create an extended region of destructive interference in the image plane, called a \emph{dark hole}, in which one may hope to detect a planet.\cite{GiveonKern_EFC11,Kasdin_EFC16b,  Desai2024_EFClab}
In test bed settings, such procedures yield dark holes with contrasts of roughly $10^{-9}$.\cite{Belikov_LabDemo_SPIE22, Seo_JPLhiContrastResult_JATIS19}
Any exoplanet light will be incoherent with the starlight. This incoherence is critical to the detection, as the central region of the exoplanet’s image will not be significantly modulated by the DM, provided the DM modulation does not appreciably degrade the Strehl ratio.\cite{Bottom_CDI_MNRAS17, Gladysz10}

Adding challenge to the situation are the nefarious effects of instrumental polarization, which arise due to reflection and refraction. 
These effects on the polarization state of the light are called \emph{polarization aberration}.\cite{McGuire90, Breckinridge15, ashcraft2024SCoOB, Ashcraft2025_GSMT2}
In this article, the fields and the corresponding intensities that exist due to polarization aberration will be referred to as \emph{secondary} components, as opposed to the \emph{primary} components that are treated when polarization aberration is ignored.
Baudoz \emph{et al.} argued that polarization aberration had an effect of $\sim10^{-8}$ on the TDH2 bench.\cite{Baudoz_Goos-Hanchen_Imbert-Fedorov}
Further, polarization dependent aberrations were part of a comprehensive study on the effect of aberrations in the Roman Space Telescope Coronagraph, indicating the polarization effects contribute about $3\times10^{-10}$ to the intensity (contrast units).\cite{Krist_End2End_Roman_JATIS23}

The secondary fields are commonly referred to as ``incoherent,"\cite{breckinridge2003polarization, guyon2009WFC_PIAA, Kasdin_EFC16, ashcraft2024SCoOB} which not technically correct (see below), but other works use the term more carefully.\cite{Breckinridge_SPIE18}
While this parlance is understandable since the secondary intensity is only weakly modulated in the vicinity of a dark hole (which is confirmed by the simulations contained herein as well), the term is not in accordance with a rigorous definition and may be misleading.
The reality is that the secondary fields are fully coherent with the primary fields, as shown later.
The adjective "incoherent'' tends to imply that the comportment of the secondary fields is indistinguishable from that of the truly incoherent planetary emission, which is not the case.\cite{Breckinridge_SPIE18}
This coherence results in correlated behavior of the various non-interfering intensity components, potentially leading to concomitant reduction of them all, even though the control (or optimization) loop seeks only to minimize the primary intensity.
We refer to this phenomenon as \emph{gratis mitigation.}  

The discussion begins with definitions of the terms ``coherent" and ``incoherent."
Next, these terms are applied to develop Jones formalism that represents the electric fields in an optical system.
The elements of the Jones matrices that represent a telescope/coronagraph system are populated using state-of-the-art physical optics simulations of the end-to-end system, resulting in a linear models of the primary and secondary fields.
In these simulations, a Lyot coronagraph is fed by a beam resulting from $4\,$m-class primary mirror that is subsequently subjected to a $\sim f/5$ beam reduction system that includes two fold mirrors, resulting in non-negligible polarization aberration in the detector plane.
We note that Lyot coronagraphs are not considered to be a robust design choice for HWO concepts.\cite{Anche_PolAb_2023, Ashcraft_coatings_JATIS25}

Example dark holes are created via minimizing either just one of, or the sum of, the two primary intensity components, yet, the components that are not included in the optimization objective are reduced as well.
The Jones matrices that represent the end-to-end optical system are subjected to a cross SVD analysis to gain insight in the gratis mitigation phenomenon.

\section{Coherence Properties of the Primary and Secondary Fields}\label{sec: Coherence}


\subsection{Stochastic Definitions of ``Coherent" and ``Incoherent"}

The theoretical treatment of polarization and interference rests on the formalism of stochastic processes, of which we avail ourselves only the simplest elements.
The propagation of electric fields through many optical systems, including stellar coronagraphs, can be carried out one frequency (or wavelength) at a time, but accounting for polarization and interference phenomena requires a bit more care, which leads to the notion of \emph{quasi-monochromatic} sources.\cite{StatisticalOptics, Collett93}
A quasi-monochromatic electric field in the $\hat{\bx}$-direction (along the $x$ axis), centered on the frequency $\nu$ may be represented as:
\begin{equation}
\bE(t) = \hat{\bx} E_x a(t) \exp[j 2 \pi \nu t] \, ,
\label{eq: def quasi-monochromatic}
\end{equation}
where $E_x$ is a complex-valued constant, representing the amplitude and phase, and  
$a(t)$ is a complex-valued stochastic process, called an \emph{envelope function} that provides rapid modulation on a time-scale that is orders-of-magnitude smaller than any conceivable detector integration time.
Eq.~\eqref{eq: def quasi-monochromatic} is a good representation of an electric field that results from passing light arising from thermal source, such as a star, through a narrow-band filter.

For simplicity, consider an unpolarized beam with fields in the $\hat{\bx}$ and $\hat{\by}$ directions.
These fields in the two orthogonal directions are specified by statistically independent envelope functions, $a(t)$ and $b(t)$, respectively.
Similarly to Eq.~\eqref{eq: def quasi-monochromatic}, the vector electric field can be represented as:
\begin{equation}
\bE(t) = \big[\hat{\bx} E_x a(t) + \hat{\by} E_y b(t) \big]\exp(j 2 \pi \nu t) 
\, ,
\label{eq: unpolarized quasi-monochromatic}
\end{equation}
where $b(t)$ is another envelope function (see below),  $E_x$ and $E_y$ are complex-valued constants (with $|E_x|=|E_y|$ since the beam is unpolarized).
To generalize Eq.~\eqref{eq: unpolarized quasi-monochromatic} to a beam with arbitrary Stokes parameters $(I,\, Q, \, U, \, V)$, one can specify the fields in the $x$ and $y$ directions via:  $E_x(t) = A a(t)$, and $E_y(t) = (\exp{j\phi})[ Ba(t) + Cb(t)]$ (where the harmonic term $\exp(j 2 \pi \nu t)$ has been dropped), and then solve for the real-valued constants $A, \, B, \, C$ and $\phi$ using the relations: $I = \overline{|E_x|^2} + \overline{|E_y|^2}$, $Q = \overline{|E_x|^2} - \overline{|E_y|^2}$, $U=2\Re(\overline{E_xE_y^*})$, $V=-2\Im(\overline{E_xE_y^*})$ (increasing phase convention), \cite{Collett93, Born&Wolf} where the superscript $^*$ indicates complex conjugation and the overbar indicates a time-average operator, which is equivalent to taking the mean of a stochastic process for the purposes of this article.

The statistically independent envelope functions $a(t)$ and $b(t)$ are
subject to the following conditions:\cite{Collett93, StatisticalOptics}
\begin{align}
& \mathcal{P}[a(t),b(t' + \tau)]  = \mathcal{P}[a(t)] \mathcal{P}[b(t' + \tau)] \; , \forall \tau \label{eq: stat indep} \\
& \overline{a(t)}  = \overline{b(t)}  = 0. \label{eq: zero mean envelopes} \\
& \overline{a(t)a^*(t)} = \overline{b(t)b^*(t)} = 1 \label{eq: unity variance} \\
& \overline{a(t)b(t + \tau)}  = \overline{a(t)b^*(t + \tau)}  = \overline{a^*(t)b(t + \tau)} = 0 \; , \forall \tau  \label{eq: incoherent envelopes}
\, ,
\end{align}
where the $\mathcal{P}$ represents probability.   
In reality, integration over a finite amount of time is needed for Eqs.~\eqref{eq: zero mean envelopes} through~\eqref{eq: incoherent envelopes} to be effectively realized.
We assume that the required integration times are much less than any currently possible detector frame rate (say, $10^{-6}\,$s).
Eq.~\eqref{eq: stat indep} states that the processes $a(t)$ and $b(t)$ are statistically independent.
Eq.~\eqref{eq: zero mean envelopes} states that the envelope functions are zero-mean, and 
Eq.~\eqref{eq: unity variance} states that the envelope functions have a variance of unity.
Eq.~\eqref{eq: incoherent envelopes} defines \emph{incoherence} as the vanishing of all first-order cross-correlations between the two envelope functions. 
Eq.~\eqref{eq: incoherent envelopes}, in fact, follows from Eqs.~\eqref{eq: stat indep} and Eq.~\eqref{eq: zero mean envelopes}.
Eq.~\eqref{eq: incoherent envelopes} is the definition of ``incoherence,'' and no other will suffice.
Eq.~\eqref{eq: unity variance} indicates that the functions $a(t)$ and $b(t)$ are both fully self-coherent.
Two envelope functions $a(t)$ and $c(t)$ with unity variance are \emph{fully coherent} if $\big|\overline{a(t)c^*(t + \tau)}\big| = 1 $ for some value of $\tau$, and \emph{partially coherent} if 
$0 < \mathrm{max}_\tau \big|\overline{a(t)c^*(t + \tau)} \big| < 1 \,$.\cite{StatisticalOptics}

\subsection{Stochastic Fields and Their Intensities}

When an unpolarized beam encounters an optical system that introduces no differential delays comparable to or greater than the coherence time $\tau_\mathrm{c}$, defined as the smallest value of $\tau_\mathrm{c}$ such that  $\overline{|a(t)a^*(t + \tau)|} \lesssim 1/2 $, then field in the output plane is given by:\cite{Breckinridge_SPIE18}
\begin{equation}
\begin{pmatrix}
E_x'(t) \\
E_y'(t)
\end{pmatrix}
=
\begin{pmatrix}
J_{xx} & J_{yx} \\
J_{xy} & J_{yy}
\end{pmatrix}
\begin{pmatrix}
	E_x a(t) \\
	E_y b(t)
\end{pmatrix}  \, ,
\label{eq: Jones stochastic}
\end{equation}
where the harmonic term $\exp(j 2 \pi \nu t)$ has been amputated.  
The $2\times 2$ matrix containing complex valued quantities $ J_{xx}, \, J_{xy} , \,
J_{yx},\,\mathrm{and} \; J_{yy}$ is the celebrated Jones Matrix.\cite{IntroFourierOptics}
Carrying out the matrix-vector multiplication in Eq.~\eqref{eq: Jones stochastic} results in the following vector field in the output plane:
\begin{equation}
\bE'(t) = \underbrace{\hat{\bx}J_{xx} E_x a(t) + \hat{\by}J_{yy} E_y b(t)   }_{\text{primary fields}}  +
  \underbrace{\hat{\bx} J_{yx}E_y b(t)   + \hat{\by} J_{xy} E_x a(t)  }_{\text{secondary fields}} \, ,
\label{eq: Jones stochastic output}
\end{equation}
\begin{wrapfigure}{l}{0.45\textwidth}
	\hspace{-5mm}
	\includegraphics[width=0.52\textwidth]{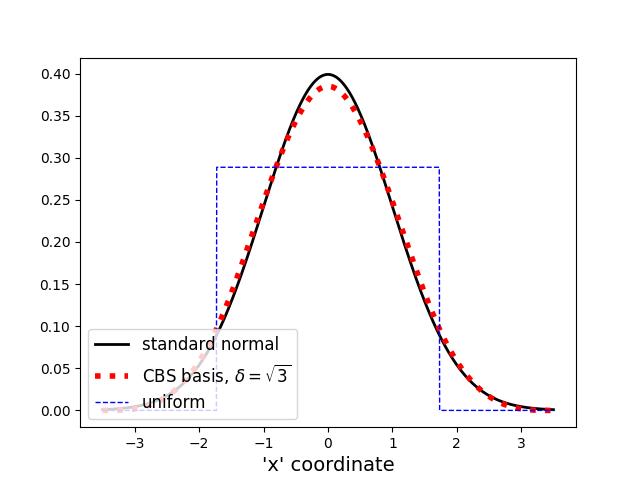}
	\caption{\small A cardinal B-spline (CBS) basis function in 1D with the width parameter $\delta$ set  to $\sqrt{3}$ and the standard normal probability density function for comparison.  The uniform (pixel) basis functions is shown in addition.  All three functions integrate to unity and have a variance of unity as well.}\label{fig: CBS1D}
	\vspace{-15mm}
\end{wrapfigure}
where the primary and secondary fields have been identified.
In Eq.~\eqref{eq: Jones stochastic output}, it is self-evident that the primary field component $\hat{\bx}J_{xx} E_x a(t)$ and the secondary field component $\hat{\by} J_{xy} E_x a(t) $ are, in fact, \emph{fully coherent} with each other since their time dependence is governed by the same function, $a(t)$, in addition to both having the same factor $E_x$.   
Of course, the same sentiment applies to the primary field component $\hat{\by}J_{yy} E_y b(t)$ and the secondary field component $\hat{\bx} J_{yx} E_y b(t) $.

The intensity in the detector plane corresponding to Eq.~\eqref{eq: Jones stochastic output} is:
\begin{align}
I & = \overline{\bE'(t) \cdot \bE'^*(t)}   \nonumber \\
& = \underbrace{ |J_{xx} E_x |^2 + |J_{yy} E_y |^2  }_{\text{primary intensity}}  +
\underbrace{|J_{xy}E_x|^2   +  |J_{yx} E_y|^2  }_{\text{secondary intensity}} \, , \label{eq: detector intensity} 
\end{align}
where $\cdot$ represents the scalar (i.e., dot) product, and Eq.~\eqref{eq: unity variance} has been exploited.
Notice that squaring and then taking the time average of the four terms in Eq.~\eqref{eq: Jones stochastic output} results in only four terms in Eq.~\eqref{eq: detector intensity}.
This  is due to the fact that $\hat{\bx} \cdot \hat{\by}=0$ and the incoherence conditions in Eq.~\eqref{eq: incoherent envelopes}.
The above discussion shows that coherence does not imply interference.

{\bf Technical Detail:}  When treating an arbitrary input beam specified by Stokes parameters $(I,\, Q, \, U, \, V)$, as explained briefly after Eq.~\eqref{eq: unpolarized quasi-monochromatic}, the input $E_y(t)$ in Eq.~\eqref{eq: Jones stochastic} may have a second term containing the envelope function $a(t)$, which provides the needed non-zero correlation between inputs $E_x(t)$ and $E_y(t)$.
This will result in new terms additional to the four shown in Eq.~\eqref{eq: detector intensity}.  This does not pose an issue for this article because the light incident on the primary mirror is assumed unpolarized and the simulations underlying the linear models begin at the primary mirror.  If, instead, the simulations began at a later point in the optical system where polarization aberration is already present, it would be necessary to generalize the analysis.

\section{Linear Model Formalism}\label{sec: Linear Model}

This section presents the linear formalism, based on the Jones calculus, that effectively captures the behavior of the end-to-end model of the telescope/coronagraph system simulated in this article. 
Under the linear model formalism, the complex-valued scalars $ J_{xx}, \, J_{yy} , \, J_{xy},\,\mathrm{and} \; J_{yx}$ in Eq.~\eqref{eq: Jones stochastic} become complex-valued matrices that map the DM phasor, which is a 2D, complex-valued function, to the electric fields at locations of the pixels in the coronagraph detector. 
These four Jones matrices comprise the ``matrix-based model."
The value of the matrix-based model is twofold: 
\begin{enumerate}
	\item{It provides a basis for a linear analysis that explains the correlation of the various intensity components at the origin of the gratis mitigation phenomenon (see Sec.~\ref{sec: cross spectrum}, below).}
	\item{Once the Jones matrices have been constructed, the evaluation of DM commands is fast on a standard CPU, only requiring lightweight computations to represent the phasor corresponding to the DM command and then one matrix-vector multiply per Jones matrix. (A Python implementation on the author's outdated desktop computer takes about $0.2\, s$ to calculate the fields in the detector plane resulting from a DM command.) }
\end{enumerate}

\subsection{Representation of the DM Phasor}\label{sec: CBS}

In this analysis, the DM phasor (see below for a precise definition) is represented by a vector with complex-valued entries, in which each entry corresponds to the coefficient of a \emph{cubic cardinal B-spline} (CBS) basis function.
The CBS basis functions perform a spatial decomposition of the coronagraph entrance pupil, effectively ``pixelizing" it.
The CBS basis functions are commonly used in image processing and optics for interpolation of continuous functions.
The Fourier transform of the CBS basis function has sidelobes that are orders-of-magnitude smaller than those of the square pixel basis function, which makes it advantageous for optical propagation.\cite{UnserSpline}  
For reference, the CBS basis function, has a width parameter $\delta$, and is given by the formula:
\begin{equation}
\eta(x; \delta) = 
\begin{cases}
0, & \text{if } \left|\dfrac{x}{\delta}\right| \geq 2 \\
\dfrac{(2 - |x/\delta|)^3}{6}, & \text{if } 1 \leq \left|\dfrac{x}{\delta}\right| < 2 \\
\dfrac{2}{3} - \left(\dfrac{x}{\delta}\right)^2 + \dfrac{1}{2}\left(\dfrac{x}{\delta}\right)^3, & \text{if } 0 \leq \left|\dfrac{x}{\delta}\right| < 1 \, .
\end{cases}
\label{eq: CBS 1D}
\end{equation}
Thus, $\eta(x; \delta)$ has compact support (i.e., is nonzero) only in the interval $(-2 \delta, 2 \delta)$.
The CBS basis functions, separated by the distance $\delta$, overlap with their neighbors and are not orthogonal.
The noteworthy properties of the CBS function include the following identities for its $0$\underline{th} and $2$\underline{nd} moments:
\begin{equation}
\frac{1}{\delta}\int_{-2 \delta}^{2\delta} \eta(x;\delta) \mathrm{d} x  =  1 \, \: \: \: \mathrm{and} \: \: 
\frac{1}{\delta}\int_{-2 \delta}^{2\delta} \eta(x;\delta)x^2 \mathrm{d} x  =  \frac{\delta^2}{3}  \,.
\label{eq: Unser moments}
\end{equation}
The function $(1/\sqrt{3})  \eta(x; \sqrt{3})$ bears a striking resemblance to the univariate standard normal probability density function (PDF), as depicted in Fig.~\ref{fig: CBS1D}, which also includes the corresponding uniform PDF.
The two dimensional (2D) version of this CBS basis function is simply the tensor product of the one-dimensional function with itself; so, the 2D version is:
\begin{equation}
\eta(x,y;\delta) = \eta(x;\delta) \eta(y;\delta) \, .
\label{eq: CBS 2D}
\end{equation}

The process of translating the real-valued DM command vector into the needed vector of CBS coefficients representing the DM phasor has three steps:
\begin{enumerate}
	\item{The height of the DM's membrane is modeled as a CBS expansion, which interpolates the DM command vector.  This results in a real-valued, essentially continuous 2D function, called the ``height function."  Other interpolation schemes, such as summing Gaussians,\cite{Kasdin_EFC16b} resulting in the height function could serve as well.  In these simulations, the DM command vector, denoted as $\bc$, has 441 components, corresponding to a $21 \times 21$ array of actuators. }
	\item{Assuming normal incidence, the real-valued height function is multiplied by $j4\pi / \lambda$ (where $\lambda$ is the wavelength and $j = \sqrt{-1}$) and placed inside the exponential function to convert it into a phasor.  The result is a continuous, complex-valued 2D function, called ``the DM phasor."}
	\item{To formulate a matrix-based model, the DM phasor must represented in terms of basis functions that tile the coronagraph entrance pupil.  Whereas the Fourier transform of a square pixel basis function has large sidelobes, the CBS basis functions have sidelobes that are orders-of-magnitude smaller, making them a reasonable choice.\cite{Frazin_CoherenceDarkHole}
	Interpolation of the DM phasor onto the CBS basis functions via linear least-squares regression results in a vector of complex-valued coefficients, here denoted as $\ba$.
	This expansion has a denser grid of CBS basis functions than does the expansion for the DM height function.  The denser grid provides robustness to the nonlinearties that grow as the size of the gradients in the DM phasor increase.  In other words, there is no linearization in phase of the DM phasor, but accurate approximation of the DM phasor requires a sufficiently dense set of basis functions.  In these simulations, this second spline grid is $33\times33$ to support the $21 \times 21$ grid of actuators on the DM.  The set of $33^2=1089$ complex-valued coefficients is denoted by the vector $\ba$.  See the Technical Detail below.}
\end{enumerate}
Thus, the CBS coefficient vector $\ba$ is a nonlinear function of the DM command vector $\bc$, and we represent this functional relationship with the notation $\ba(\bc)$, but $\ba$ should be understood to be a function of $\bc$ when this functional dependence is  not explicit in the notation.
One advantage of this scheme is that all nonlinearity in the model is confined to the function $\ba(\bc)$.

{\bf Technical Detail:}  While the DM phasor has a magnitude of unity, the interpolation of the CBS coefficients fit to said DM phasor may depart from this ideal when the gradient of the DM phasor is too large.\cite{Frazin_CoherenceDarkHole}
To mitigate this problem, the dark hole optimization loop (see Sec.~\ref{sec: results}) was constrained to not allow neighboring DM actuators to correspond to a phase difference greater than $\pi/2$.   
The DM commands corresponding to the four dark holes resulted in 4 DM phasors all with very similar statistics: a mean amplitude departing from unity by about $\pm 5\times10^{-4}$, and a standard deviation of about $7\times10^{-3}$.   
While this may not be sufficiently accurate for a calibrated computational model (although comparison to standard linearization schemes should be made before this conclusion can be drawn), it is sufficient for this article's goal of documenting the gratis mitigation phenomenon.

\subsection{Jones System Matrices}\label{sec: Jones Matrices}

This section presents the Jones model of the end-to-end optical system.
While the quantities $ J_{xx}, \, J_{yy} , \, J_{xy},\,\mathrm{and} \; J_{yx}$ in Eq.~\eqref{eq: Jones stochastic} are complex-valued scalars, their analogues that describe the vector-field response of the optical system to the DM phasor are complex-valued matrices.
The matrices $\bD_{xx}$ and $\bD_{xy}$ represent the responses to $x-$polarized light incident on the primary mirror, and $\bD_{yy}$ and $\bD_{yx}$ represent the responses to $y-$polarized light incident on the primary mirror.
\newpage

\begin{wraptable}[27]{l}{0.52\textwidth}
	\hspace{-1mm}
	\begin{tabular}{|l|c|}
		\hline
		Pre-Coronagraph Optics & \\
		\hline
		wavelength ($\lambda$) & $0.9 \, \mu$m \\
		primary mirror width ($W$) & 4.24 m \\
		primary mirror focal length  & 24 m \\
		dist. to mirror \#1 & 8 m \\
		mirror \# 1 tilt angle & $20^\circ$ \\
		dist. to mirror \#2 & 8 m \\
		mirror \# 2 tilt angle & $-20^\circ$ \\ 
		dist. to collimating OAP & 8.11 m \\
		collimating OAP focal length & 110 mm \\
		collimating OAP off-axis angle & $10^\circ$ \\
		\hline 
		Coronagraph Optics & \\
		\hline
		coronagraph entrance pupil ($w$) & 19.34 mm \\
		coronagraph effective  $f$\# & 44 \\
		DM actuators (spanning $w$) & $21 \times 21$ \\
		occulter opaque width & $142 \, \mu$m \\
		occulter transiton edge width & $71 \, \mu$m \\
		inner working angle & $ \sim 3.1 \, \lambda/w$ \\
		Lyot stop clear width & $9.1 \,$mm \\
		Lyot stop transition edge width & $1.2 \,$mm \\
		OAP1 focal length ($f$) & 800 mm \\
		OAP2, OAP3 focal length ($f$) & 400 mm \\
		OAP1, OAP2, OAP3  & $20^\circ$ \\
		\hspace{4mm} off-axis angle ($\zeta$)& \\
		\hline
	\end{tabular}
	\caption{\small System parameters used for the simulations.  The primary mirror, occulter and Lyot stop are all square.  The other mirrors are circular and large enough to contain the square beam.  All mirrors are protected silver.}
	\label{table: optical system}
\end{wraptable}
These matrices are $M\times N$, where $M$ is the number of detector pixels and $N$ is the number of spline coefficients used to specify the DM phasor as outlined in Sec.~\ref{sec: CBS}.
Column $n$ of any these four ``$\bD$" matrices corresponds to the field in the detector plane when the coronagraph entrance pupil is multiplied by a mask corresponding to the $n$\underline{th} CBS basis function.  

Following Sec.~\ref{sec: CBS}, a DM command vector $\bc$ results in vector of spline coefficients $\ba$.
Then, the vector field in the detector plane is given by a generalization of Eq.~\eqref{eq: Jones stochastic output} to the system model. For an unpolarized input beam impinging on the primary mirror, this is:
\begin{align}
\bE(t) &= \underbrace{\hat{\bx}\bD_{xx} \ba a(t) + \hat{\by}\bD_{yy} \ba b(t)   }_{\text{primary fields}}    \nonumber \\
& + \underbrace{\hat{\bx} \bD_{xy} \ba a(t)   + \hat{\by} \bD_{yx} \ba b(t)  }_{\text{secondary fields}} \, ,
\label{eq: Jones stochastic output big}
\end{align}
where the vector $\bE(t)$ has length $M$, but has both $x$ and $y$ directions, formally making it a vector of 2-element vectors. (Fortunately, this equation is only for pedagogical, not computational, purposes.)   
Following logic that is similar to that used to go from Eq.~\eqref{eq: Jones stochastic output} to Eq.~\eqref{eq: detector intensity}, we find that the intensity corresponding to Eq.~\eqref{eq: Jones stochastic output big} is:
\begin{align}
I & = \overline{\bE(t) \cdot \bE^*(t)}   \label{eq: detector intensity 0} \\
& = \underbrace{ |\bD_{xx} \ba |^2 + |\bD_{yy} \ba |^2  }_{\text{primary intensity}}  +
\underbrace{|\bD_{xy} \ba|^2   +  |\bD_{yx} \ba|^2  }_{\text{secondary intensity}} \, . \label{eq: detector intensity big}
\end{align}
Thus, the matrices $\bD_{xx}$ and $\bD_{yy}$ provide the primary intensities, while $\bD_{yx}$ and $\bD_{xy}$ provide the secondary intensities.  
Below, in Sec.~\ref{sec: results}, it will be seen that, when viewed as functions of $\ba$ (which itself is a function of the DM command $\bc$) ,$|\bD_{xx} \ba |^2$ and $|\bD_{yy} \ba |^2$ have similar behavior, as do $|\bD_{xy} \ba|^2$ and $|\bD_{yx} \ba |^2$, but there are differences that merit attention.

\section{The Telescope/Coronagraph Model}\label{sec: end-to-end}

\begin{figure}[t]
	\centering
	\includegraphics[width=0.8\textwidth]{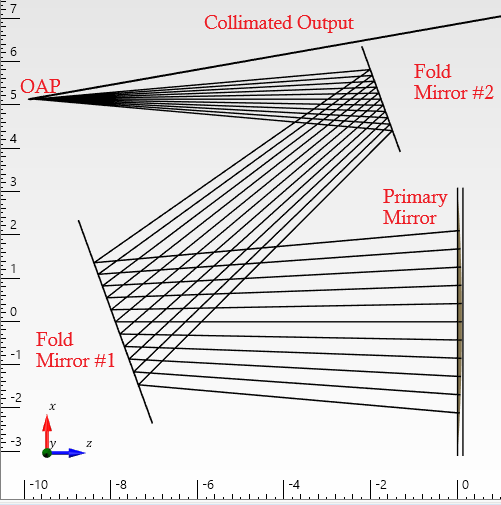}
	\caption{\small A schematic diagram of the pre-coronagraph optics resulting in the collimated output beam.  The $y$ axis is invariant in this system. Note that the rays from the star impinging on the primary mirror are not displayed for clarity.  The units of the rulers are meters.  See Fig.~\ref{fig: BigBeam_detail} for detail on the collimation scheme.  The coronagraph, not shown here, can be seen in Fig.~\ref{fig: coronagraph}.}
	\label{fig: BigBeam_full}
\end{figure}

State-of-the-art physical optics simulations undergird the construction of the $\bD_{xx}, $ $\bD_{yy}$, $\bD_{xy}$, and $\bD_{yx}$ matrices.
Briefly put, $x-$polarized light is incident on the primary mirror to create $\bD_{xx}$ and $\bD_{xy}$ simultaneously.
First, the light is propagated through the pre-coronagraph optical train shown in Fig.~\ref{fig: BigBeam_full}, whereupon it is multiplied by mask corresponding to one of the CBS basis functions in the coronagraph entrance pupil.
Subsequently, the light is passed through the coronagraph, shown in Fig.~\ref{fig: coronagraph}, to the detector plane, where the $x$ and $y$ components of the field are recorded. 

The entire optical design is symmetric about the $x-z$ plane.
Thus, the coordinate transformations used by the simulator to evaluate quantities in any particular plane are rotations about the $y$ axis.
This is what is meant by ``$y$-axis invariance."
As a result of this invariance, $x-$polarized light incident on the primary mirror results in primary and secondary components in the detector plane that are $x-$ and $y-$polarized, respectively.
Similarly, $y-$polarized light incident on the primary mirror results in fields that are $y-$ and $x-$polarized for the primay and secondary components, respectively.

\begin{figure}[t]
	\hspace{-12mm}
	\begin{tabular}{r l}
		\includegraphics[width=0.55\textwidth]{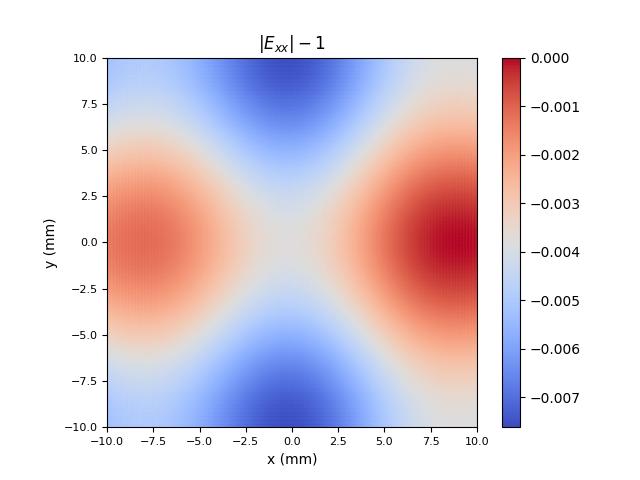}	&
		\includegraphics[width=0.55\textwidth]{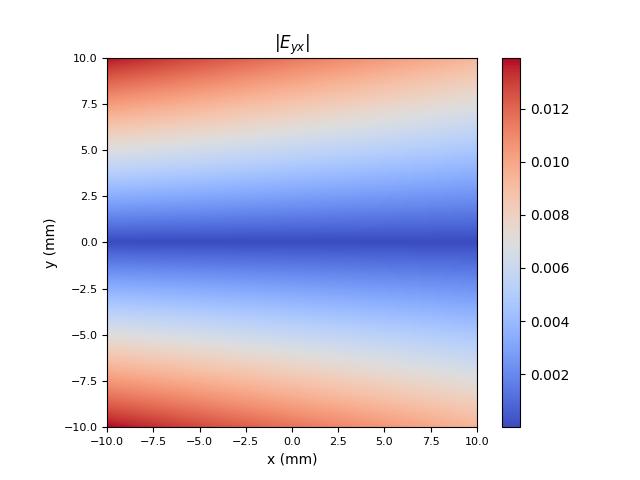} \\
		\includegraphics[width=0.55\textwidth]{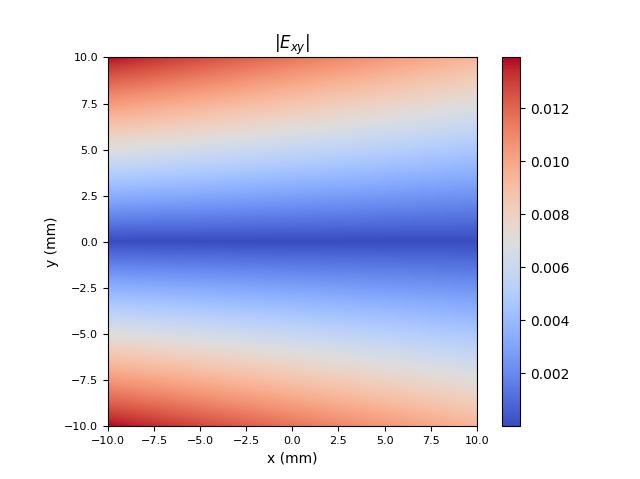}	&
		\includegraphics[width=0.55\textwidth]{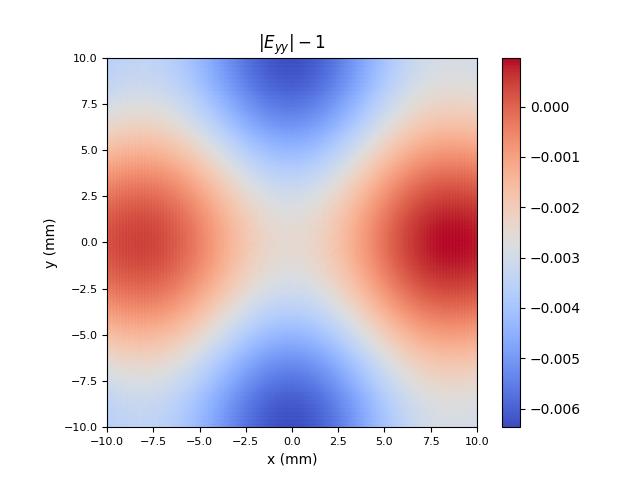} 
	\end{tabular}
	\caption{\small Depiction of the Jones pupil at the coronagraph entrance on a linear color scale. \emph{Upper Left:} $|E_{xx}|-1$.   \emph{Upper Right:} $|E_{yx}|$. \emph{Lower Left:} $|E_{xy}|$.  \emph{Lower Right:} $|E_{yy}|-1$.  The fields are normalized so that the maximum value of $|E_{xx}|$ is unity.  While the images of $|E_{yx}|$ and $|E_{xy}|$ are visually indistinguishable, close examination reveals subtle differences between the images of $|E_{xx}|-1$ and $|E_{yy}|-1$.}
	\label{fig: Pupil Fields}
\end{figure}

Fig.~\ref{fig: Pupil Fields} is a depiction of the Jones pupil at the coronagraph entrance, which evinces the effects of polarization aberration of the beam reduction optics.
For the purposes of this figure, all of the fields are normalized by the maximum value of $|E_{xx}|$.
Since the magnitudes of the primary components $|E_{xx}|$ and $|E_{yy}|$ are nearly uniform, the displayed quantities on the diagonal are $|E_{xx}|-1$ and $|E_{yy}|-1$.
$|E_{xx}|-1$ and $|E_{yy}|-1$ are similar to each other, as are $|E_{xy}|$ and $|E_{yx}|$.

\begin{figure}[]
	\centering
	\includegraphics[]{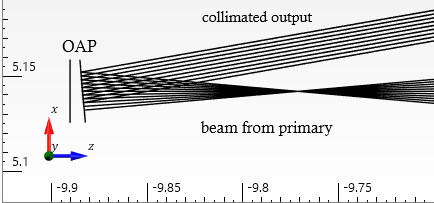}
	\caption{\small Detail of Fig.~\ref{fig: BigBeam_full} displaying the recollimation via an OAP.  The ``collimated output" in this diagram is the ``input beam" in Fig.~\ref{fig: coronagraph}.  The units of the rulers are meters.}
	\label{fig: BigBeam_detail}
\end{figure}

\begin{figure}[t]
	\centering
	\includegraphics[width=1.1\textwidth]{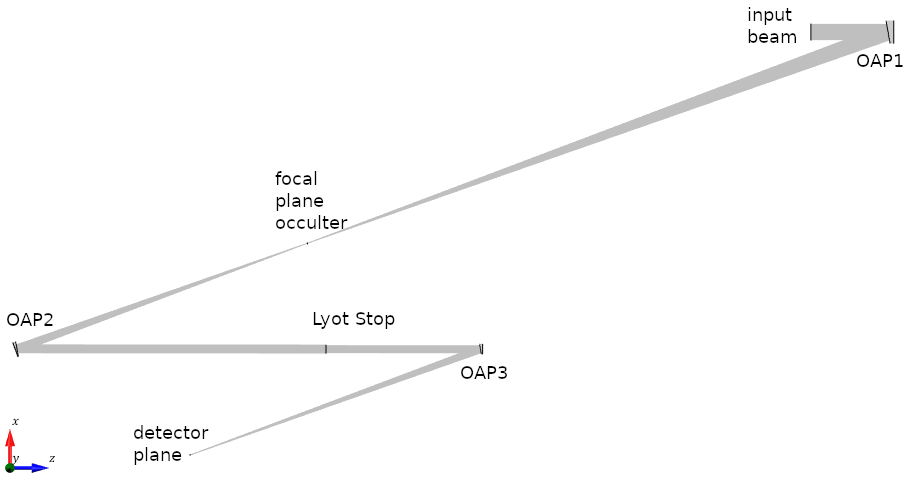}
	\caption{\small A schematic diagram of the Lyot-type stellar coronagraph used in these simulations. The partially polarized input beam is provided by the optical system shown in Figs.~\ref{fig: BigBeam_full} and \ref{fig: BigBeam_detail}. 
		The DM, which is not shown, modulates the beam before it reaches OAP1. The $y$ axis is invariant in this system.}
	\label{fig: coronagraph}
\end{figure}

\begin{figure}[t]
	\hspace{-5mm}
	\begin{tabular}{r l}
		\includegraphics[width=0.5\textwidth]{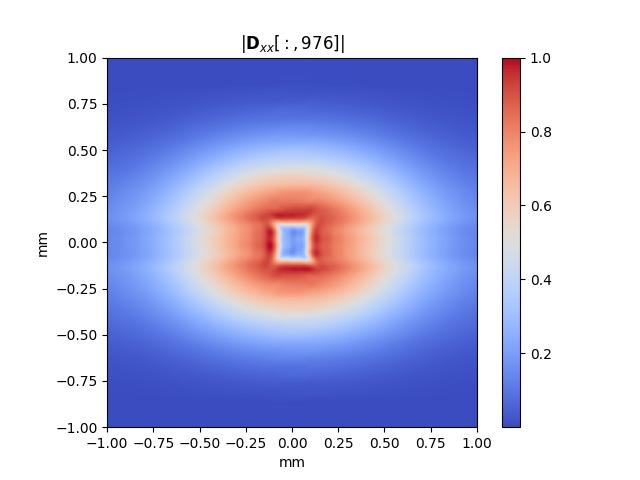}	&
		\includegraphics[width=0.5\textwidth]{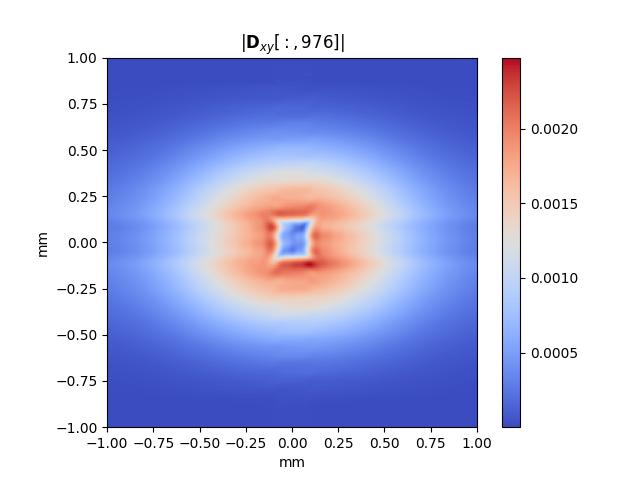} \\
		\includegraphics[width=0.5\textwidth]{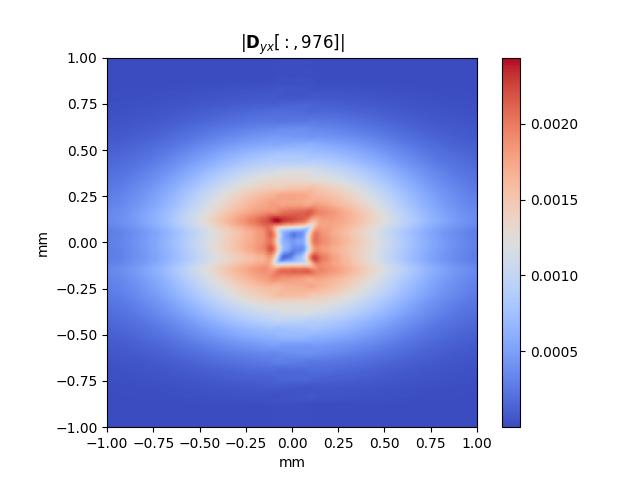}	&
		\includegraphics[width=0.5\textwidth]{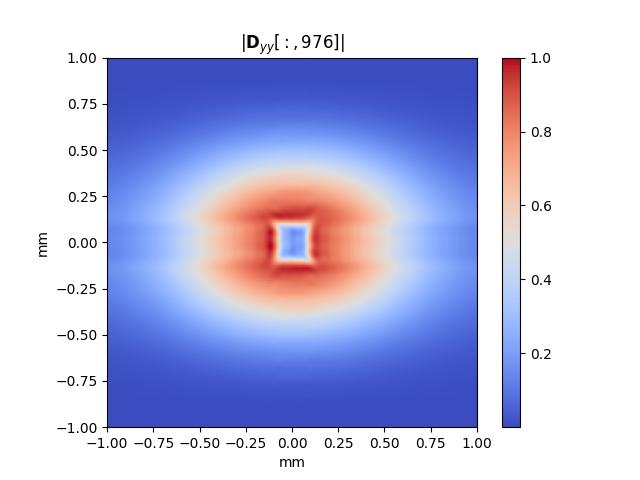} 
	\end{tabular}
	\caption{\small Absolute values of detector plane fields associated with the CBS basis function centered at $(29\delta,\, 19\delta)$ (corresponding to linear index 976), where $\delta$ is the grid spacing of the spline knots in the entrance pupil.  Thus, the displayed quantities are the absolute values of column 976 of the matrices $\bD_{xx}$ (\emph{upper left}), $\bD_{yy}$ (\emph{lower right}),  $\bD_{xy}$ (\emph{upper right}).  The color scale is linear.  Note that these values include the polarization aberration in the coronagraph entrance pupil seen in Fig.~\ref{fig: Pupil Fields}.  The dark square in the center is due to the coronagraph's occulter.}\label{fig: Jones spline}
\end{figure}

\clearpage

\begin{wrapfigure}[20]{l}{0.45\textwidth}
	\hspace{0mm}
	\includegraphics[width=0.45\textwidth]{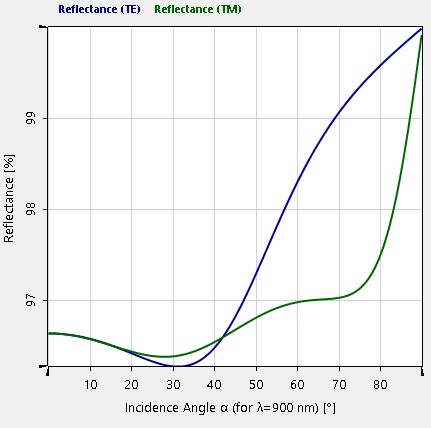}
	\caption{\small Magnitudes of the complex-valued TM and TE reflection coefficients at $\lambda=900\,$nm as a function of the angle of incidence for a silver film and a $470 \,$nm thick SiO$_2$ coating.}\label{fig: OAP coating}
	\vspace{0mm}
\end{wrapfigure}
Fig.~\ref{fig: Jones spline} shows the magnitudes of the fields in the coronagraph detector plane that result from illuminating only CBS basis function \#976, which is at the position $(29, \, 19)$ on the $33 \times 33$ grid of basis functions in the coronagraph entrance pupil.  The dark spot in the center is a result of the occulter.  For the diagonal (primary) components, the structure is due to diffraction in the coronagraph.  
The anti-diagonal (secondary) components would not exist without polarization aberration and they have additional structure due to diffraction and the occulter.

\subsection{Optical Layout}\label{sec: Layout}

\begin{figure}[ht]
	\hspace{-11mm}
	\begin{tabular}{r l}
		\hspace{-4mm}
		\includegraphics[width=.61\textwidth]{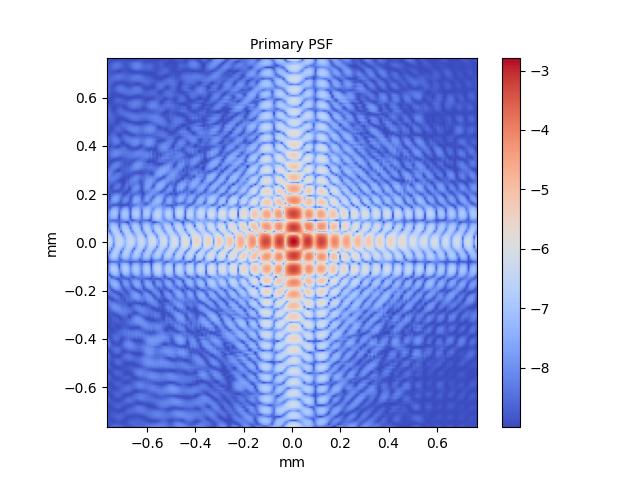} & 
		\hspace{-19mm}
		\includegraphics[width=.61\textwidth]{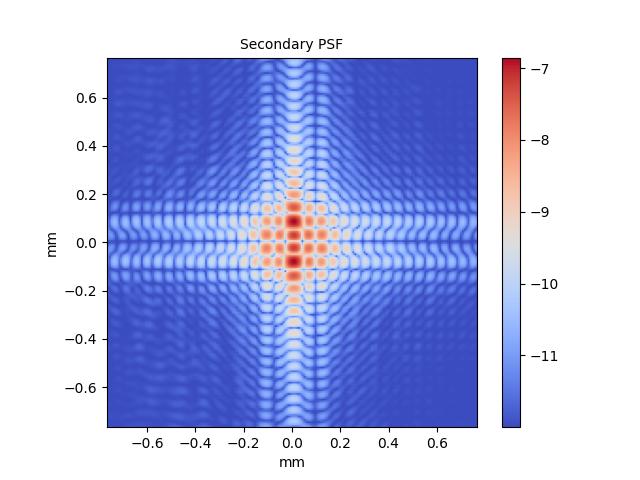}
	\end{tabular}
	\caption{\small Nominal PSFs corresponding to the end-to-end simulations.
		\emph{Left:}  Primary $I_{xx}$  PSF (contrast units).   
		\emph{Right:}  Secondary $I_{xy}$ PSF (contrast units). 
		The $I_{yy}$ and $I_{yx}$ PSFs are visually indistinguishable from the two PSFs shown here.  See also Fig.~\ref{fig: All Holes}.
	}
	\label{fig: PSFs}
\end{figure}

This portion describes the optical system in the simulations.
The layout, in which the beam from the primary mirror feeds a Lyot coronagraph, is summarized in Table~\ref{table: optical system}, and Figs.~\ref{fig: BigBeam_full}, \ref{fig: BigBeam_detail} and \ref{fig: coronagraph}.
In order to illuminate all of the DM actuators, which are taken to be mounted in a square array with $21\times21$ elements, these simulations use a square beam.
The input beam is incident on the square primary mirror with a width ($W$) of $4.24\,$m, which makes it $6 \,$m on the diagonal.
The square optical elements are the primary mirror,  the coronagraph's entrance pupil, the coronagraph's occulter and its Lyot stop.
The OAPs and the fold mirrors, are circular and sized to not clip the square beam.      
The primary mirror is an on-axis parabola with a focal length of $24\,$m, which makes it $f/4$ on the diagonal and $f/5.65$ in the $x$ and $y$ directions.  
This coronagraph consists of 3 OAPs, an occulter in the initial focal plane, and a Lyot stop in the plane corresponding to Fourier transform of the field in the initial focal plane.
The square geometry is plainly evident in the PSFs provided in Fig.~\ref{fig: PSFs}.

\section{Physical Optics}\label{sec: prop methods}

While optical system design is simple, the propagation methods are state-of-the-art.
The propagation engine is provided by the VirtualLab Fusion (VLF) optical simulation package developed by LightTrans, GmbH.
VLF is a comprehensive physical optics software platform widely used in photonics and microscopy [e.g., \citenum{Wyrowski_DiffMetaLens_SPIE19,  Wyrowski_NanoOpt_JOSAA2020}].
It supports a broad range of algorithms for free-space propagation and surface interaction, enabling self-consistent treatment of vectorial, geometrical, and diffractive phenomena.
VLF's algorithms do not rely on the assumption of paraxiality, which is implicit in the Fresnel and Fraunhofer approximations in Fourier optics,\cite{IntroFourierOptics} unless explicitly chosen by the user.
VLF always represents the electromagnetic field itself without ever resorting to a ray-based representation.
Indeed, the ``rays"  shown in Figs.~\ref{fig: BigBeam_full}, \ref{fig: BigBeam_detail}, and \ref{fig: coronagraph} are not rays at all-- they are lines connecting sets of points on input and output surfaces (either of which can be curved) that are determined via a mapping algorithm.
This surface-to-surface mapping is a central component of VLF's field-based geometrical propagation algorithms.  

While VLF's geometrical propagation algorithms explicitly exclude diffraction,  VLF's non-geometrical algorithms incorporate diffraction via the angular spectrum method.\cite{IntroFourierOptics}
Since both the geometrical and non-geometrical algorithms utilize a field-based representation, applying them in sequence to different stages of the propagation presents no difficulties.  
For example, if one stage of a propagation is carried out in the non-geometrical mode, and is followed by a second stage carried out in the geometrical mode, the fringes incurred due to diffraction in the first stage will be inherited and treated self-consistently by the geometrical propagation in the second stage, so the diffractive structure from the first stage will not be destroyed.\cite{Wyrowski_PolarizFieldTr_SPIE2019, Wyrowski_Seamless_SPIE24, Frazin_CoherenceDarkHole}
The reverse is true as well, i.e., one can follow a geometrical propagation, in which there are no diffractive effects, with non-geometrical propagation that imposes diffractive structure on the field.
In fact, this is precisely the strategy used in these simulations: All of the propagation steps from the primary mirror to the coronagraph entrance pupil are carried out in geometrical mode, which was well-suited to determine the polarization aberration imposed by the pre-coronagraph optical train (see Fig.~\ref{fig: Pupil Fields}).
Starting at the coronagraph entrance pupil, all of the propagation steps were carried  out in non-geometrical mode to account for diffraction through the coronagraph up to the detector.

VLF employs the Local Plane Wave Interface Approximation (LPIA) to compute the interaction of the vector electromagnetic field with dielectric multilayers, such as protected silver mirrors.\cite{Wyrowski_LPIA_AplOpt00}
At each point $p$ on the mirror surface, the LPIA determines the local direction of propagation of the incident field, which is defined as the direction of the phase gradient at $p$.
The electric field vector is then locally decomposed into transverse electric (TE) and transverse magnetic (TM) components, relative to the local plane of incidence at $p$.
Fig.~\ref{fig: OAP coating} shows the magnitudes of the TM and TE reflection coefficients as a function of angle for the OAPs in this simulation.
The reflected field vector is then computed using the Transfer Matrix Method (TMM), applied to the multilayer stack under the local angle of incidence.\cite{Born&Wolf}
The resulting complex-valued TE and TM reflection coefficients are then used to construct the reflected field vector.
A detailed discussion of VLF's treatment of an OAP mirror is provided in Ref.~[\citenum{Frazin_CoherenceDarkHole}].

\section{Results}\label{sec: results}


This section is devoted to providing examples of the simulations and an explanation of the dark hole optimizations.
Eq.~\eqref{eq: detector intensity big} shows that we must consider two primary intensity components $|\bD_{xx} \bpac |^2 $ and $|\bD_{yy} \bpac |^2 $, and two secondary components $|\bD_{xy} \bpac |^2 $ and  $|\bD_{yx} \bpac |^2 $ in coronagraph detector.
This notation emphasizes the fact that these four components are functions of the DM command $\bc$, however, it will prove less tedious to use shorthand notation, in which these same quantities are denoted as $I_{xx}$, $I_{yy}$, $I_{xy}$ and $I_{yx}$, respectively.

Fig.~\ref{fig: PSFs} provides the nominal (i.e., flat DM) primary $I_{xx}$ and secondary $I_{yy}$ PSFs in contrast units.
The corresponding images of the other primary and secondary PSFs, i.e.  $I_{yy}$ and  $I_{yx}$, are visually indistinguishable from the two shown here and, hence, are not provided, but closeups of all four are shown in the top row of Fig.~\ref{fig: All Holes}.  

The contrast normalization was taken to be the primary intensity of an $x-$polarized source that is off-axis by an angle of $5 \lambda/w$ (where $w$ is the width of the square coronagraph entrance pupil), which makes the effects of the coronagraph's occulter negligible.
This off-axis source is simulated by fitting the complex-valued spline coefficient vector $\ba$ to the 2D phasor corresponding to a unit amplitude and the linear phase with a spatial frequency of $5/w$.
This is well within Nyquist frequency of the $33\times33$ array of CBS basis functions, which supports spatial frequencies up to about $14/w$ (see Sec.~\ref{sec: CBS}).
Note that is procedure bypasses the DM by specifying the vector $\ba$ directly, so in this one instance $\ba$ is not a function of the DM command $\bc$ (but a non-zero $\bc$ can be used to modify this off-axis PSF, if desired).
Since the effects of polarization aberration are negligible for purposes of determining the contrast normalization factor, this procedure is equivalent to an off-axis beam impinging on the primary mirror at an off-axis angle of $5 \lambda/W$.
  
\begin{figure}[h!]
	\hspace{-8mm}
	\begin{tabular}{r l}
		\includegraphics[width=.55\textwidth,height=.50\textwidth]{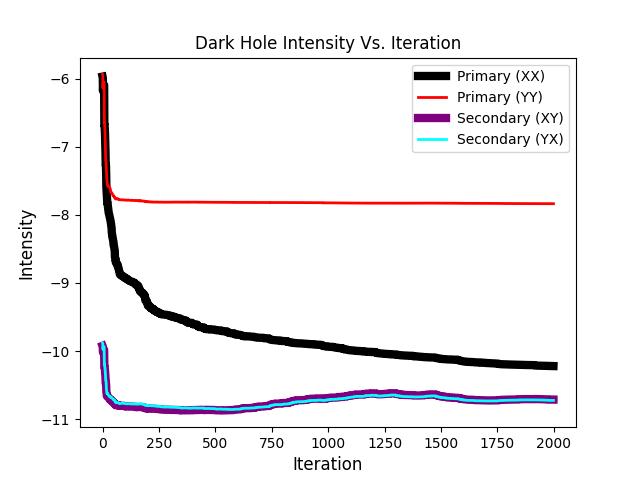} & \hspace{-10mm}
		\includegraphics[width=.55\textwidth,height=.50\textwidth]{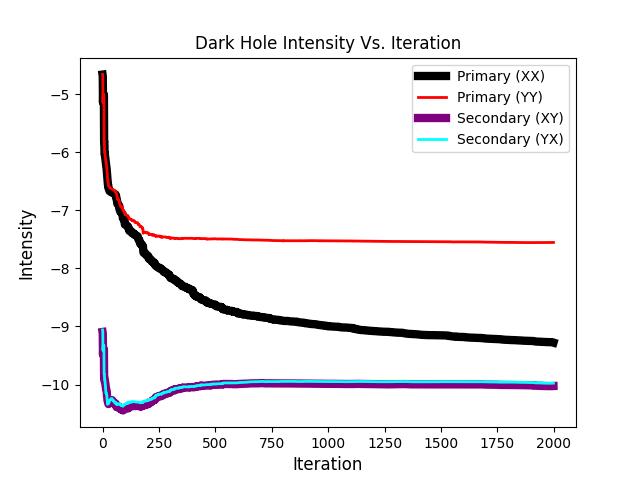} \\ 
		\includegraphics[width=.55\textwidth,height=.50\textwidth]{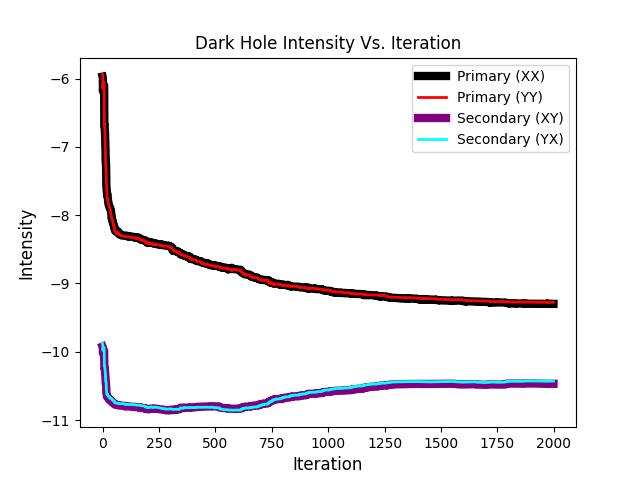} & \hspace{-10mm}
		\includegraphics[width=.55\textwidth,height=.50\textwidth]{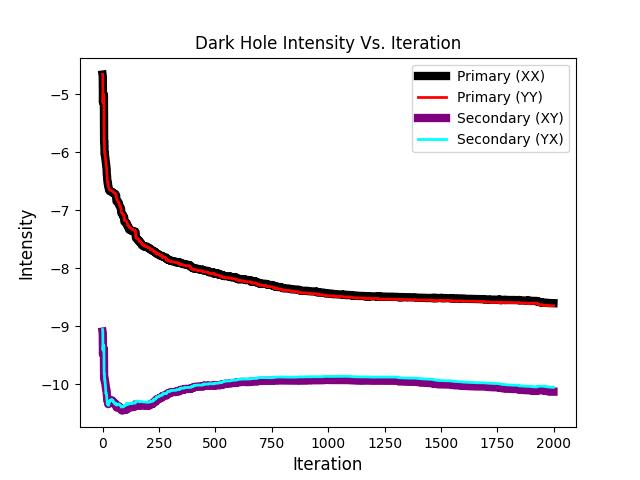}
	\end{tabular}
	\caption{Spatially averaged intensities (contrast units) in dark holes \#1 and \#2 as a function of the constrained conjugate gradient (CG) iteration number.  The initial iteration corresponds to the nominal, i.e., flat-DM, state.
		\emph{Left panels:} Dark hole \#1.  \emph{Right panels:} Dark hole \#2.
		\emph{Upper Panels:} Minimization of only the primary intensity $I_{xx}$.
		\emph{Lower panels:} Minimization of the the sum of the two primary intensities, i.e., $I_{xx} + I_{yy}$.  In all panels, the black, red, purple and light blue correspond to $I_{xx}$, $I_{yy}$, $I_{xy}$, and $I_{yx}$, respectively, and the total intensity is the sum of all four terms.  See also Table~\ref{table: intensities}.  The dark holes displayed in  Fig.~\ref{fig: All Holes} correspond the final iterations shown here.
	}
	\label{fig: DH iterations}
\end{figure}

\subsection{The Dark Holes}

This dark hole procedure is unlike most EFC methods in that there is no need of a sensing step due to the fact that the electric field is a known function of the DM command for these models.
Thus, optimization of the intensity, which is a form of ``energy minimization,"\cite{Malbet_EFC95} is all that is required to dig dark holes in this specific context.
Here we present the results of four different dark hole optimizations that correspond to two separate locations in the detector plane and two different optimization objectives for each location.  
In order to ensure the CBS representation of the DM phasor has an amplitude near unity (see Sec.~\ref{sec: CBS}), all optimizations are constrained so that phases of neighboring DM actuators do not differ by more than $\pi/2$. 
The first objective function for the optimization is the spatial mean of one of the primary intensity components, $I_{xx}$, over the dark hole pixels.
The second objective function is the spatial mean of the sum of the two primary intensity components, i.e., $I_{xx} + I_{yy}$.
All optimizations are performed with respect to the DM command vector $\bc$, which, in turn, specifies the coefficient vector $\ba$, as explained in Sec.~\ref{sec: CBS}.

\begin{figure}[h!]
	\hspace{-1mm}\vspace{-3mm}
	\includegraphics[width=0.95\textwidth]{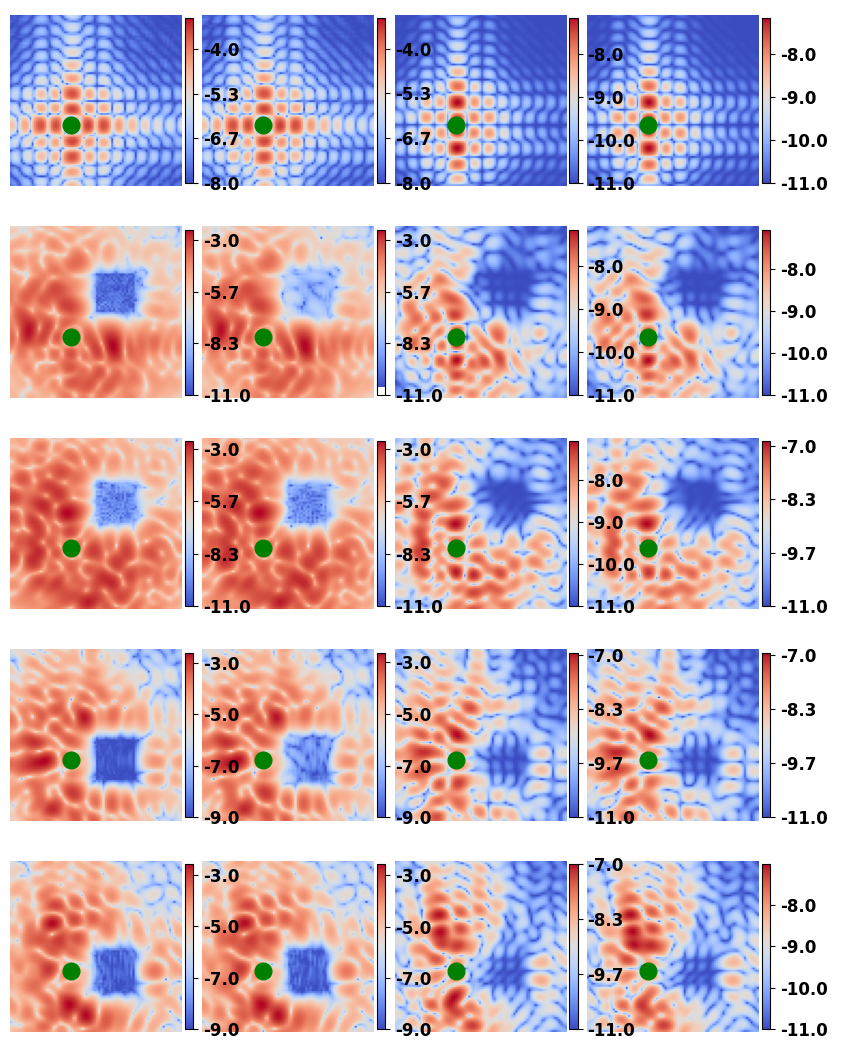}
	\caption{
		$\mathrm{Log}_{10}$-intensities in a portion of the coronagraph detector plane slightly upward and to the right of center, which encompasses the origin (green dot) as well as dark holes \#1 and \#2.  Each row corresponds to different DM solution, and the total intensity is the sum of all four images of a given row.  For display purposes, it is not possible to place all of the images on a common scale.  The color scale corresponding to each image is on its right.
		\emph{Top Row:} Nominal configuration (flat DM).
		\emph{2\underline{nd} Row:} min$\left[I_{xx} \right]$ for dark hole \#1.  \emph{3\underline{rd} Row:} min$\left[I_{xx} + I_{yy}\right]$ for dark hole \#1.  \emph{4\underline{th} Row:} min$\left[I_{xx} \right]$ for dark hole \#2. 
		\emph{5\underline{th} Row:} min$\left[I_{xx} + I_{yy} \right]$ for dark hole \#2.
		\emph{Columns 1-4} show $I_{xx}$, $I_{yy}$, $I_{xy}$, and $I_{yx}$, respectively.
		The mean intensities over the dark hole regions can be seen in Fig.~\ref{fig: DH iterations} and Table~\ref{table: intensities}.
	} 
	\label{fig: All Holes}
\end{figure}

Both dark holes are square. 
Dark hole \#1 is centered on the $45^\circ$ line in the image plane, and it extends from 2 to 5 $\lambda/w$ in both the $x$ and $y$ directions, where $w$ is the width of the square coronagraph entrance aperture. (Due to pre-coronagraph optics, the angle $s/w$ in the coronagraph, where $s/w << 1$, projects to an angle of about $(s/W)$ on the sky.)
Dark hole \#2 straddles the $x$ axis, on which it extends from 2 to 5 $\lambda/w$, and it is between -1.5 and +1.5 $\lambda/w$ in $y$.
Fig.~\ref{fig: DH iterations} provides the spatial means of the intensities of the dark holes as a function of constrained conjugate gradient iteration number for all four optimizations, and Fig.~\ref{fig: All Holes} shows the resulting (final iteration) dark holes, which are easily seen in all four intensity components.

\clearpage

The gratis mitigation phenomenon, in which some optimization process with the objective of minimizing one (or more) the intensity components also results in concomitant reduction of other intensity components, is clearly evident in Figs.~\ref{fig: DH iterations} and~\ref{fig: All Holes}.
Consider the second and fourth rows in Fig.~\ref{fig: DH iterations}.
Both of these are the result of minimizing of only $I_{xx}$ over the dark hole region.  In both cases, the minimization reduces $I_{xx}$ by about 4 orders-of-magnitude, yet, it also serves to reduce $I_{yy}$ by about 2 two orders-of-magnitude.
Additionally, $I_{xy}$ and $I_{yx}$ are reduced by roughly one order-of-magnitude (see also Table~\ref{table: intensities}).
The results of minimizing $I_{xx} + I_{yy}$ over the dark hole regions are displayed in the third and fifth rows.
Under this optimization objective, $I_{xx}$ and  $I_{yy}$ are both reduced by the same 3 orders-of-magnitude.
The resulting dark holes can be seen in the third and fifth rows of Fig.~\ref{fig: All Holes}.  
The secondary intensity components, $I_{xy}$ and $I_{yx}$ behave rather similarly with either optimization objective, with roughly an order-of-magnitude of gratis mitigation.

It is worth taking a moment to analyze the behavior of the $I_{xy}$ and $I_{yx}$ secondary intensities in more detail.
Inspection of Fig.~\ref{fig: DH iterations} shows that they do exhibit the monotonic behavior of the primary intensities and obtain their minima, which are about 60\% smaller than the final iteration values, early in the iteration process (these particular values are provided in the caption of Table~\ref{table: intensities}).
Since the optimization process does not take $I_{xy}$ and $I_{yx}$ into account, there is no reason they should exhibit monotonic behavior.
Referring to Table~\ref{table: intensities}, the gratis mitigation of the secondary intensities is a significant effect in these examples.
Using the $^0$ and $^\mathrm{f}$ superscripts to represent the flat DM and final iteration values, respectively, we make several observations:
\begin{itemize}
\item Dark hole \#1, min($I_{xx}$): $I_{yx}^0 / I_{xx}^\mathrm{f} \approx 2 $ and $I_{yx}^\mathrm{f} / I_{xx}^\mathrm{f} \approx 0.3 $.
In this case, the gratis mitigation would be of paramount importance if such an measurement were made with a linear polarizer just before the detector to remove the $I_{yy}$ and $I_{xy}$ components.
\item Dark hole \#1, min($I_{xx} + I_{yy}$):  $(I_{xy}^0 + I_{yx}^0) / (I_{xx}^\mathrm{f} +I_{yy}^\mathrm{f} ) \approx 0.25 $ and  $(I_{xy}^f + I_{yx}^f) / (I_{xx}^\mathrm{f} +I_{yy}^\mathrm{f} ) \approx 0.07 $.
\item Dark hole \#2, min($I_{xx}$): $I_{yx}^0 / I_{xx}^\mathrm{f} \approx 1.5 $ and $I_{yx}^\mathrm{f} / I_{xx}^\mathrm{f} \approx 0.2$.
Again, the gratis mitigation would be of paramount importance in this example if such an measurement were made with a linear polarizer just before the detector to remove the $I_{yy}$ and $I_{xy}$ components.
\item Dark hole \#2, min($I_{xx} + I_{yy}$):  $(I_{xy}^0 + I_{yx}^0) / (I_{xx}^\mathrm{f} +I_{yy}^\mathrm{f} ) \approx 0.34 $ and  $(I_{xy}^f + I_{yx}^f) / (I_{xx}^\mathrm{f} +I_{yy}^\mathrm{f} ) \approx 0.04 $.  Thus, gratis mitigation plays a critical role here.

\end{itemize}
In all of these cases, attempting disentangle planetary light from the starlight near the contrast limit of the dark holes would require accounting for gratis mitigation.

\begin{figure}[h!]
	\includegraphics[width=0.75\textwidth]{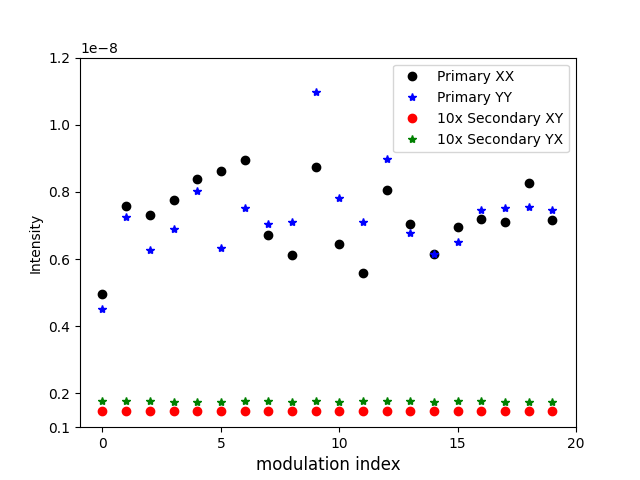}
	\caption{Modulation with random DM perturbations about the solution corresponding to the final iteration in the lower right panel of Fig.~\ref{fig: DH iterations}.  The random phases are spatially white and from a standard normal distribution with a standard deviation of $5\times10^{-4}$ radians.  The $y$ axis is the intensity in contrast units.  The $x$ axis index of the random DM command, with index 0 corresponding to perturbation of zero (which yield the dark hole itself). Both primary intensities and secondary intensities are shown, the latter of which are multiplied by 10 for display.}
	\label{fig: modulation}
\end{figure}

\subsection{Modulation}
Fig.~\ref{fig: modulation} shows the results of applying random perturbations to the DM command corresponding to the dark hole solution from the final iteration displayed in the lower right panel of Fig.~\ref{fig: DH iterations}.
The random phases are spatially white and from a standard normal distribution with a standard deviation of $5\times10^{-4}$ radians.
Recall that this solution is the result of minimizing the spatial mean of $I_{xx} + I_{yy}$ over the pixels in dark hole \#2, so $I_{xx}$ and $I_{yy}$ are minimized with nearly equal efficacy here.
The figure shows the dark hole solution itself at index 0, and indices 1-19 correspond to random perturbations. 
These random perturbations result in intensities with standard deviations relative to the dark hole values (index 0) of 23\%, 19\%, 0.4\% and 0.4\% for $I_{xx}$, $I_{yy}$, $I_{xy}$, and $I_{yx}$ respectively.
It is  not surprising that $I_{xx}$ and $I_{yy}$ are much more sensitive to these perturbations than $I_{xy}$, and $I_{yx}$ because the dark hole command is much closer to local minima of  $I_{xx}$ and $I_{yy}$ than it is to local minima of $I_{xy}$, and $I_{yx}$.   
This illustrates the difficulty of modulating intensity components that are not near their extrema for a given dark hole command.

\subsection{Spectral Analysis}\label{sec: cross spectrum}

This section takes advantage of the linear model formalism to provide insight into the gratis mitigation phenomenon, which is fundamentally a result of the similities in the responses to the DM of the four Jones matrices.
First, take $q$ to be a vector of 1D detector pixel indices corresponding to the dark hole under consideration, and then let $\bD_{xx}^q$, $\bD_{yy}^q$, $\bD_{xy}^q$ and $\bD_{yx}^q$ be versions of their namesakes that only include the row indices in $q$ (no columns are removed).  
These system matrices correspond only to the dark hole pixels.
Let $\bW_1$ be one of $\bD_{xx}^q$, $\bD_{yy}^q$, $\bD_{xy}^q$ or $\bD_{yx}^q$, and let $\bW_2$ be a different one.

Consider the singular value decompositions (SVDs) of $\bW_1$ and $\bW_2$.\cite{Moon&Stirling}
These SVDs are denoted as $\bU_1  \bSigma_1 \bV_1^* = \bW_1$ and $\bU_2  \bSigma_2 \bV_2^* = \bW_2$, where $^*$ is the Hermitian conjugate operator.
$\bU_1$, $\bU_2$, $\bV_1$, and $\bV_2$ are unitary, complex-valued matrices, and the $\bSigma_1$ and $\bSigma_2$ are diagonal matrices containing the singular values, which are non-negative.
The vectors corresponding to the the singular values along the diagonals of $\bSigma_1$ and $\bSigma_2$ are sorted into descending order, as is customary. 
This vector of singular values is also called the \emph{spectrum} of a matrix.  
This analysis will exploit the following property of the SVD: $|\bW_1 \bv_{1k} |_2 = \sigma_{1k} $, where $|\,|_2$ represents the 2-norm, and $\bv_{1k}$ is column $k$ of the matrix $\bV_1$, also referred as a \emph{right singular vector}.
Thus, the $k$\underline{th} singular value can be obtained by multiplying the original matrix by its its  $k$\underline{th} singular vector and taking the 2-norm of the result.
Doing this for all of the right singular vectors results in the spectrum. 

The \emph{cross spectra} of two matrices are close relatives of their spectra, and they provide a method of comparing these matrices that emphasizes their dominant modes.
First, recall that the best (in a least-squares sense) rank$-N$ approximation of the matrix $\bW_1$ is $\sum_{k=0}^{N-1}  \sigma_{1k}\bu_{1k}\bv_{1k}^* $, where $\bu_{1k}$ is the $k$\underline{th} left singular vector of $\bW_1$ and $\bu_{1k}\bv_{1k}^*$ is an outer (tensor) product of the two vectors.  
Thus, it can be seen that the singular value is an indicator of the importance of a given mode.  
For two matrices $\bW_1$ and $\bW_2$ with matching dimensions, the magnitude of the response of $\bW_2$ to mode $k$ of $\bW_1$ is  $(1/\sigma_{20})|\bW_2 \bv_{1k}|_2$, where the factor $(1/\sigma_{20})$ provides normalization by dividing out the largest singular value of $\bW_2$.   
Carrying out this calculation over all of the modes results in a cross spectrum.

\begin{figure}[ht]
	\hspace{-8mm}
	\begin{tabular}{r l}
		\includegraphics[width=0.51\textwidth]{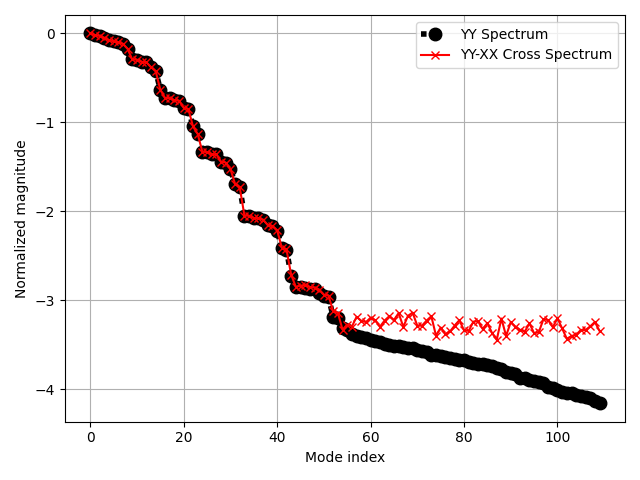}	&
		\includegraphics[width=0.51\textwidth]{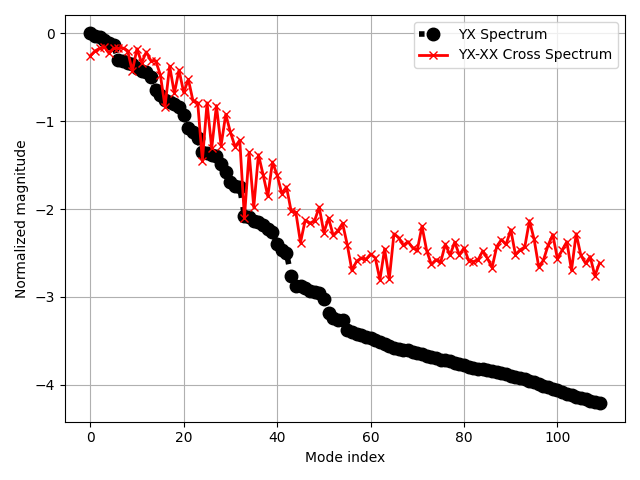} \\
		\includegraphics[width=0.51\textwidth]{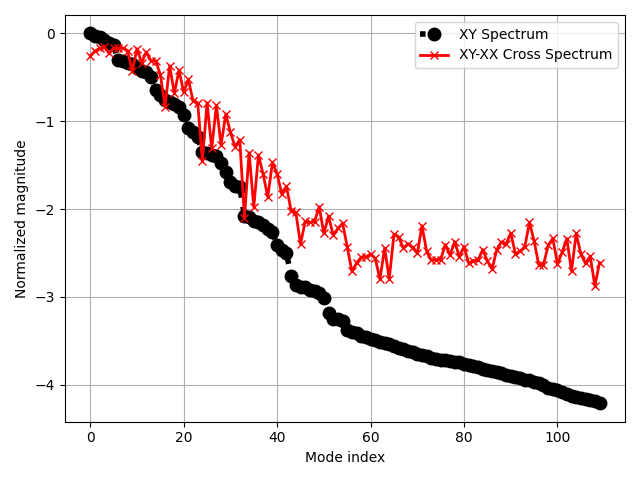}	&
		\includegraphics[width=0.51\textwidth]{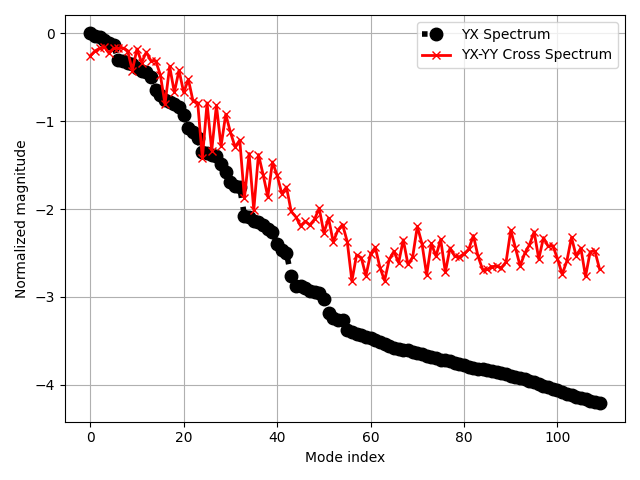} 
	\end{tabular}
	\caption{Spectra and cross spectra for the dark hole \#1 pixels (denoted by $q$).   
	\emph{Upper Left:} Spectrum of $\bD_{yy}^q$ (black) and the cross spectrum from multiplying $\bD_{yy}^q$ by the right singular vectors of $\bD_{xx}^q$. The discord measure, defined in Eq.~\eqref{eq: diff metric}, for these curves is $\mathcal{D}=8.4$.
	\emph{Upper Right:} Spectrum of $\bD_{yx}^q$ (black) and the cross spectrum from multiplying $\bD_{yx}^q$ by the right singular vectors of $\bD_{xx}^q$.  $\mathcal{D}=24.7$.
	\emph{Lower Right:} Spectrum of $\bD_{yx}^q$ (black) and the cross spectrum from multiplying $\bD_{yx}^q$ by the right singular vectors of   $\bD_{yy}^q$.  $\mathcal{D}=24.1$.
	\emph{Lower Left:} Spectrum of $\bD_{xy}^q$ (black) and the cross spectrum from multiplying $\bD_{xy}^q$ by the right singular vectors of   $\bD_{xx}^q$.  $\mathcal{D}=24.5$.  All curves are normalized by the largest singular value as explained in the text.
   } \label{fig: cross spectra}
\end{figure}

Based on Figs.~\ref{fig: DH iterations}, \ref{fig: All Holes}, and~\ref{fig: modulation}, it is fair to say that the behaviors of the two primary intensities, $I_{xx}$ and $I_{yy}$, are much more similar to each other than they are to the behaviors of the secondary intensities, $I_{xy}$ and $I_{yx}$, and vice-versa.
The spectra and cross spectra displayed in Fig.~\ref{fig: cross spectra}, are consistent with this observation.
To be quantitative, we can define a \emph{discord measure} for a spectrum and a cross spectrum, represented by the vectors $\bg$ and $\bs$, respectively, in which $\bg$ is assumed to be normalized by its largest value, via the formula:
\begin{equation}
\mathcal{D}(\bs,\bg) = \sqrt{ \sum_{n=0}^{N-1} \big[\log(s_n) - \log(g_n) \big]^2   } \, , 
\label{eq: diff metric}
\end{equation}
where $N$ is the number of modes deemed to be significant, and $s_n$ and $g_n$ are the $n$\underline{th} components of the vectors $\bs$ and $\bg$, respectively.
For all four matrices, i.e., $\bD_{xx}^q$, etc., and both dark holes, less than 1 part in $10^8$ of the power (in terms of the squared Frobenius norm) of these matrices resides in all of the modes with indices beyond 110, thus $N$ is taken to be 110 in these examples.  

Fig.~\ref{fig: cross spectra} exhibits spectra and cross spectra in which the pixels in $q$ correspond to dark hole \#1.   
Consider the upper left panel of Fig.~\ref{fig: DH iterations}, in which only $I_{xx}$ was optimized to create the dark hole, and $I_{yy}$ undergoes 2 orders of gratis mitigation.
To understand this instance of gratis mitigation, it is helpful to determine the response of the matrix $\bD_{yy}^q$ to the dominant modes of $\bD_{xx}^q$.
The upper left panel of Fig.~\ref{fig: cross spectra} shows the spectrum of $\bD_{yy}^q$  and its cross spectrum with singular vectors of $\bD_{xx}^q$, which we will call the ``YY-XX cross spectrum," with the ``YY" coming before the ``XX" in this nomenclature  because the cross spectrum components are calculated by multiplying the $\bD_{yy}^q$ matrix on the right by the singular vectors of $\bD_{xx}^q$.
The roles of $\bD_{xx}^q$ and $\bD_{yy}^q$ are reversed in the XX-YY cross spectrum.
The XX-YY cross spectrum is not displayed because the response of $\bD_{xx}^q$ to the modes of $\bD_{yy}^q$ does little to explain the gratis mitigation of $I_{yy}$ when $I_{xx}$ is minimized.
The goal of the other three panels of Fig.~\ref{fig: cross spectra} is to provide insight into the gratis mitigation of $I_{xy}$ and $I_{yx}$ when either $I_{xx}$ or $I_{xx}+I_{yy}$ are minimized, so similar logic underlies the choice to display YX-XX, XY-XX, and YX-YY cross spectra in these panels.

The upper left panel of Fig.~\ref{fig: DH iterations} demonstrates two orders-of-magnitude of gratis mitigation of $I_{yy}$ (see also Table~\ref{table: intensities}) when $I_{xx}$ is minimized, where as $I_{yx}$ and $I_{xy}$ have less than one order-of-magnitude of gratis mitigation when $I_{xx}$ or $I_{xx}+I_{yy}$ are minimized.
The comparison of the spectra and cross spectra shown in Fig.~\ref{fig: cross spectra} are indicative of the gratis mitigation in each case:  The upper left figure shows very close agreement of the spectrum of $\bD_{yy}^q$ and its cross spectrum with $\bD_{xx}^q$ over the first 55 modes, and the discord measure, $\mathcal{D}$ in Eq.~\eqref{eq: diff metric}, is 8.4.    
The other three cross spectra displayed in the other panels follow the general trend of the spectra for perhaps the first 30 modes, but the correspondence is not nearly as close as it is in the upper left panel.  
Their discord measures reflect the difference in the amount of gratis mitigation, and are all about 24.

\section{Discussion} 

This article is a simulation study that is the first to consider the evolution effects of polarization aberration in a stellar coronagraph as a sequence of DM commands converges on a dark hole.
The polarization aberrations simulated in this article arise from $4\,$m-class primary mirror with 
$\sim f/5$ beam reduction optics that includes 2 fold mirrors  (see Sec.~\ref{sec: Layout}) followed by a Lyot coronagraph, with rigorous physical optics at each step in the propagation.

\begin{table}[ht!]
	\hspace{-7mm}
	\begin{tabular}{|c|c|c|c|c|c|c|}
		\hline
		&hole $\#1$& hole $\#1$    &hole $\#1$              &hole $\#2$&hole $\#2$    &hole $\#2$ \\
		Quantity &  nominal          & min($I_{xx}$) & min($I_{xx} + I_{yy}$) &nominal   &min($I_{xx}$) & min($I_{xx} + I_{yy}$) \\
		\hline
		$I_{xx}$ & $1.1\times10^{-6 }$&$6.0\times10^{-11}$&$5.0\times10^{-10}$&$2.2\times10^{-5 }$&$5.4\times10^{-10}$&$2.5\times10^{-9 }$ \\
		$I_{yy}$ & $1.1\times10^{-6 }$&$1.5\times10^{-8 }$&$5.4\times10^{-10}$&$2.2\times10^{-5 }$&$2.0\times10^{-8 }$&$2.2\times10^{-9}$ \\
		$I_{xy}$ & $1.3\times10^{-10}$&$1.9\times10^{-11}$&$3.4\times10^{-11}$&$8.1\times10^{-10}$&$9.3\times10^{-11}$&$7.0\times10^{-11}$ \\
		$I_{yx}$ & $1.3\times10^{-10}$&$1.9\times10^{-11}$&$3.7\times10^{-11}$&$8.1\times10^{-10}$&$1.1\times10^{-10}$&$8.8\times10^{-11}$ \\
		\hline
	\end{tabular}
	\vspace{1mm}
	\caption{Values from the curves in Fig.~\ref{fig: DH iterations} (contrast units).  It is easily seen in Fig.~\ref{fig: DH iterations} that minimum values of $I_{xy}$ and $I_{yx}$ are obtained well before the final iteration.  These minimum values, not shown in this table are:
		For hole \#1, under either minimization objective, the minimum values of $I_{xy}$ and $I_{yx}$ are both about $1.4\times10^{-11}$.  For dark hole \#2, the analogous quantities are both about $3.8\times10^{-11}$.  }
	\label{table: intensities}
\end{table}

To carry out this analysis, the electric fields and their corresponding intensities are classified into primary and secondary components, the latter of which exist only because of polarization aberration.  
With unpolarized light incident on the primary mirror, the intensity in the detector plane consists two primary components, $I_{xx}$ and $I_{yy}$, and two secondary components, $I_{xy}$ and $I_{yx}$, none of which interfere with each other (see Sec.~\ref{sec: Linear Model}).
While the latter two exist only because of polarization aberration, polarization aberration can also cause $I_{xx}$ and $I_{yy}$ to respond slightly differently to DM commands, as it does in the examples shown here.

The key result of this article is that the various non-interfering intensity components exhibit correlated responses to the DM commands, which may lead to a welcome phenomenon here referred to as \emph{gratis mitigation}, as it does in these simulations.
Gratis mitigation is the beneficial circumstance in which an optimization loop with the goal of finding a DM command that minimizes one of, or the sum of, the primary intensity components also reduces other intensity components. 
Two of the examples presented here had the objective of minimizing only one of the two primary components, which is achieved by 4 orders-of-magnitude, but the other primary component, concomitantly dropped by about 2 orders-of-magnitude, whereas the secondary components fell by roughly one order-of-magnitude.
The highly correlated response of the two primary components and the less correlated responses of the secondary components can be viewed in terms of the Jones system matrices that govern the fields responsible for each component.

Finding an analytical or semi-analytical expression linking the Jones system matrices to the amount of gratis mitigation one might expect is a challenging problem.
However, the correspondences of the Jones matrices' spectra with each others' cross spectra, which can be quantified with the discord measure, introduced in Eq.~\eqref{eq: diff metric}, provides some indication of how similarly a pair of components will respond to the DM (see Fig.~\ref{fig: cross spectra}).

While the iterations of the optimization loops displayed in Fig.~\ref{fig: DH iterations} provide substantial modulation of the secondary intensities, most of this modulation is found in the early iterations, long before the dark holes are fully realized.
Fig.~\ref{fig: modulation} provides an illustrative example of the difficulty in modulating the secondary intensities in the vicinity of a dark hole command, which limits the possibilities for generalizing coherent differential imaging (CDI) to account for secondary components.\cite{Bottom_CDI_MNRAS17}
This is due to the fact that the dark hole command is near a local minimum of the optimized quantity (in this case, $I_{xx} + I_{yy}$), whereas the non-optimized quantities are not near any extremum.
That said, model-based regression paired with an optimal probing scheme may be able to separate the various components.

The correlation of the responses of various intensity components to the DM pointed out in this article should be taken into account in detailed polarization calibration efforts, but perhaps the most fruitful application of the gratis mitigation concept is in coronagraph design.
Future design concepts could potentially be optimized to benefit from gratis mitigation as a way to help meet engineering constraints on polarization aberration.  
Here, a brief \emph{gedankenexperiment} may prove illustrative:  Consider a coronagraph with extremely slow optics in which the polarization aberration is negligible.
If this coronagraph were fed by a collimated, unpolarized beam that carries no polarization aberration, no secondary intensity components would arise, and the two primary intensity components would behave identically.
Next, place a planar fold mirror into the collimated beam feeding the coronagraph.
This addition would result in spatially uniform polarization at the entrance pupil, and the consequent secondary fields at the detector would differ from the primary field only by complex scale factor.
In this case, the cross spectra and spectra, would be identical, have a discord measure of zero, and complete gratis mitigation would be expected.

Thus, spatially similar (apart from a constant multiplier) polarization aberration of two components is better placed to benefit from gratis mitigation than polarization aberration that is spatially disparate.
Indeed, these simulations serve as a good example of this concept:
Fig.~\ref{fig: Pupil Fields} shows the Jones pupil at the coronagraph entrance,
in which the primary components, $|E_{xx}|$ and $|E_{yy}|$, show a similar pattern, but the pattern of the secondary components, $|E_{xy}|$ and $|E_{yx}|$, differs markedly from the pattern in the primary components.
This does much to explain the strong gratis mitigation of $I_{yy}$ seen when $I_{xx}$ is minimized, and the relatively weak gratis mitigation of$I_{xy}$ and $I_{yx}$ seen when $I_{xx}$ (or $I_{xx} + I_{yy}$) is minimized.

Following this reasoning, consider a coronagraph design problem in which the length of the optical system is fixed, and there is a choice between a slow optical system with more fold mirrors and a fast optical systems with fewer fold mirrors.
Since the latter would result in polarization aberration that is less spatially uniform than the former, it would be less well-placed to benefit from gratis mitigation than the former. 
 The major limitation of this study is that these simulations are not tailored to correspond to existing or proposed instrument designs, and quantitative predictions of the phenomena reported here will require instrument specific efforts.

\subsection*{Data Availability Statement}
The Python codes and supporting data to reproduce the results given in this article are publicly available on GitHub at \emph{https://github.com/ColdStrayPlanet/Optics/tree/master/EFC}

\subsection*{Acknowledgments}
The author would like to acknowledge Frank Wyrowski, Michael Fitzgerald, David Marx and John Kohl for helpful discussions.
This work was funded by the Heising-Simons Foundation (grant numbers: 2020-1826, 2022-3912) and 
the National Science Foundation (award number: 2308352).

\clearpage
\bibliographystyle{spiejour} 

\begin{thebibliography}{10}
	
	\bibitem{LUVOIR_model_JATIS22}
	R.~{Juanola-Parramon}, N.~T. {Zimmerman}, L.~{Pueyo}, {\em et~al.}, ``{Modeling
		and performance analysis of the LUVOIR coronagraph instrument},'' {\em
		Journal of Astronomical Telescopes, Instruments, and Systems} {\bf 8}, 034001
	(2022).
	
	\bibitem{Mennesson2024_HWOlab}
	B.~Mennesson, R.~Belikov, E.~Por, {\em et~al.}, ``Current laboratory
	performance of starlight suppression systems, and potential pathways to
	desired habitable worlds observatory exoplanet science capabilities,'' {\em
		Journal of Astronomical Telescopes, Instruments, and Systems} {\bf 10},
	035004  (2024).
	\newblock Preprint on arXiv: 2404.18036.
	
	\bibitem{Cavarroc_IdealCoronagraph06}
	C.~{Cavarroc}, A.~{Boccaletti}, P.~{Baudoz}, {\em et~al.}, ``{Fundamental
		limitations on Earth-like planet detection with extremely large
		telescopes},'' {\em Astronomy and Astrophysics} {\bf 447}, 397--403  (2006).
	
	\bibitem{Krist_End2End_Roman_JATIS23}
	J.~E. {Krist}, J.~B. {Steeves}, B.~D. {Dube}, {\em et~al.}, ``{End-to-end
		numerical modeling of the Roman Space Telescope coronagraph},'' {\em Journal
		of Astronomical Telescopes, Instruments, and Systems} {\bf 9}, 045002
	(2023).
	
	\bibitem{GiveonKern_EFC11}
	A.~{Give'on}, B.~D. {Kern}, and S.~{Shaklan}, ``{Pair-wise, deformable mirror,
		image plane-based diversity electric field estimation for high contrast
		coronagraphy},'' in {\em Techniques and Instrumentation for Detection of
		Exoplanets V},  S.~{Shaklan}, Ed., {\em Society of Photo-Optical
		Instrumentation Engineers (SPIE) Conference Series} {\bf 8151}, 815110
	(2011).
	
	\bibitem{Kasdin_EFC16b}
	T.~D. {Groff}, A.~J. {Eldorado Riggs}, B.~{Kern}, {\em et~al.}, ``{Methods and
		limitations of focal plane sensing, estimation, and control in high-contrast
		imaging},'' {\em Journal of Astronomical Telescopes, Instruments, and
		Systems} {\bf 2}, 011009  (2016).
	
	\bibitem{Desai2024_EFClab}
	N.~Desai, A.~Potier, S.~F. Redmond, {\em et~al.}, ``Comparative laboratory
	study of electric field conjugation algorithms,'' {\em Journal of
		Astronomical Telescopes, Instruments, and Systems} {\bf 10}, 035001  (2024).
	\newblock Received 4 Mar 2024; Accepted 4 Jul 2024.
	
	\bibitem{Belikov_LabDemo_SPIE22}
	R.~{Belikov}, D.~{Sirbu}, D.~{Marx}, {\em et~al.}, ``{Laboratory demonstration
		of high contrast with the PIAACMC coronagraph on an obstructed and segmented
		aperture},'' in {\em Space Telescopes and Instrumentation 2022: Optical,
		Infrared, and Millimeter Wave},  L.~E. {Coyle}, S.~{Matsuura}, and M.~D.
	{Perrin}, Eds., {\em Society of Photo-Optical Instrumentation Engineers
		(SPIE) Conference Series} {\bf 12180}, 1218025  (2022).
	
	\bibitem{Seo_JPLhiContrastResult_JATIS19}
	B.-J. Seo, K.~Patterson, K.~Balasubramanian, {\em et~al.}, ``{Testbed
		demonstration of high-contrast coronagraph imaging in search for Earth-like
		exoplanets},'' in {\em Techniques and Instrumentation for Detection of
		Exoplanets IX},  S.~B. Shaklan, Ed., {\em Journal of Astronomical Telescopes,
		Instruments, and Systems} {\bf 11117}, 599 -- 609, International Society for
	Optics and Photonics, SPIE  (2019).
	
	\bibitem{Bottom_CDI_MNRAS17}
	M.~{Bottom}, J.~K. {Wallace}, R.~D. {Bartos}, {\em et~al.}, ``{Speckle
		suppression and companion detection using coherent differential imaging},''
	{\em \mnras} {\bf 464}, 2937--2951  (2017).
	
	\bibitem{Gladysz10}
	S.~{Gladysz}, N.~{Yaitskova}, and J.~C. {Christou}, ``{Statistics of intensity
		in adaptive-optics images and their usefulness for detection and photometry
		of exoplanets},'' {\em Journal of the Optical Society of America A} {\bf 27},
	A260000--A75  (2010).
	
	\bibitem{McGuire90}
	J.~P. McGuire and R.~A. Chipman, ``Diffraction image formation in optical
	systems with polarization aberrations. i: Formulation and example,'' {\em J.
		Opt. Soc. Am. A} {\bf 7}, 1614--1626  (1990).
	
	\bibitem{Breckinridge15}
	J.~B. {Breckinridge}, W.~S.~T. {Lam}, and R.~A. {Chipman}, ``{Polarization
		Aberrations in Astronomical Telescopes: The Point Spread Function},'' {\em
		Pub. Astron. Soc. Pacific} {\bf 127}, 445--468  (2015).
	
	\bibitem{ashcraft2024SCoOB}
	J.~N. Ashcraft, E.~S. Douglas, R.~M. Anche, {\em et~al.}, ``The space
	coronagraph optical bench (scoob): 3. mueller matrix polarimetry of a
	coronagraphic exit pupil,'' {\em arXiv preprint arXiv:2406.19616}   (2024).
	
	\bibitem{Ashcraft2025_GSMT2}
	J.~N. Ashcraft, R.~M. Anche, S.~Y. Haffert, {\em et~al.}, ``Polarization
	aberrations in next‑generation giant segmented mirror telescopes (gsmts):
	Ii. influence of segment‑to‑segment coating variations on high‑contrast
	imaging and polarimetry,'' {\em Astronomy and Astrophysics} {\bf 695}, A28
	(2025).
	\newblock © The Authors 2025, Creative Commons Attribution License.
	
	\bibitem{Baudoz_Goos-Hanchen_Imbert-Fedorov}
	P.~Baudoz, C.~Desgrange, R.~Galicher, {\em et~al.}, ``{Polarization effects on
		high contrast imaging: measurements on THD2 bench},'' in {\em Space
		Telescopes and Instrumentation 2024: Optical, Infrared, and Millimeter Wave},
	L.~E. Coyle, S.~Matsuura, and M.~D. Perrin, Eds.,  {\bf 13092}, 130926L,
	International Society for Optics and Photonics, SPIE  (2024).
	
	\bibitem{breckinridge2003polarization}
	J.~B. Breckinridge and B.~R. Oppenheimer, ``Polarization effects in reflecting
	coronagraphs for white light applications in astronomy,'' {\em arXiv preprint
		astro-ph/0309399}   (2003).
	
	\bibitem{guyon2009WFC_PIAA}
	O.~Guyon, E.~Pluzhnik, F.~Martinache, {\em et~al.}, ``High contrast imaging and
	wavefront control with a piaa coronagraph: Laboratory system validation,''
	{\em arXiv preprint arXiv:0911.1307}   (2009).
	
	\bibitem{Kasdin_EFC16}
	A.~J.~E. {Riggs}, N.~J. {Kasdin}, and T.~D. {Groff}, ``{Recursive starlight and
		bias estimation for high-contrast imaging with an extended Kalman filter},''
	{\em Journal of Astronomical Telescopes, Instruments, and Systems} {\bf 2},
	011017  (2016).
	
	\bibitem{Breckinridge_SPIE18}
	J.~B. {Breckinridge}, M.~{Kupinski}, J.~{Davis}, {\em et~al.}, ``{Terrestrial
		exoplanet coronagraph image quality polarization aberrations in
		Habex},'' in {\em Space Telescopes and Instrumentation 2018:
		Optical, Infrared, and Millimeter Wave},  M.~{Lystrup}, H.~A. {MacEwen},
	G.~G. {Fazio}, {\em et~al.}, Eds., {\em Society of Photo-Optical
		Instrumentation Engineers (SPIE) Conference Series} {\bf 10698}, 106981D
	(2018).
	
	\bibitem{Anche_PolAb_2023}
	R.~M. {Anche}, S.~Y. {Haffert}, J.~N. {Ashcraft}, {\em et~al.}, ``{Estimation
		of polarization aberrations and their effect on the coronagraphic performance
		for future space telescopes},'' {\em arXiv e-prints} , arXiv:2309.04563
	(2023).
	
	\bibitem{Ashcraft_coatings_JATIS25}
	J.~N. {Ashcraft}, B.~D. {Dube}, E.~S. {Douglas}, {\em et~al.}, ``{Comparison of
		polarization aberrations from existing mirror coatings for coronagraphic
		imaging of habitable worlds},'' {\em Journal of Astronomical Telescopes,
		Instruments, and Systems} {\bf 11}, 015002  (2025).
	
	\bibitem{StatisticalOptics}
	J.~W. {Goodman}, {\em Statistical Optics, 2nd Edition}, John Wiley and Sons,
	Inc.  (2015).
	
	\bibitem{Collett93}
	E.~{Collett}, {\em Polarized Light: Fundamentals and Applications}, Marcel
	Drecker, Inc.  (1993).
	
	\bibitem{Born&Wolf}
	M.~{Born} and E.~{Wolf}, {\em Principles of Optics: Electromagnetic Theory of
		Propagation, Interference and Diffraction of Light, seventh expanded
		edition}, The Press Syndicate of the University of Cambridge  (1999).
	
	\bibitem{IntroFourierOptics}
	J.~W. {Goodman}, {\em Introduction to Fourier Optics, second edition}, The
	McGraw-Hill Companies, Inc.  (1996).
	
	\bibitem{UnserSpline}
	M.~Unser, ``Splines: a perfect fit for signal and image processing,'' {\em IEEE
		Signal Processing Magazine} {\bf 16}(6), 22--38  (1999).
	
	\bibitem{Frazin_CoherenceDarkHole}
	R.~A. {Frazin}, ``{Instrumental Polarization in Stellar Coronagraphy: Coherent
		Behavior and its Implications for Dark Hole Optimization},'' {\em arXiv
		e-prints} , arXiv:2508.00237  (2025).
	
	\bibitem{Wyrowski_DiffMetaLens_SPIE19}
	F.~Wyrowski, S.~Zhang, L.~Yang, {\em et~al.}, ``{Modeling of
		diffractive/meta-lenses using fast physical optics (Conference
		Presentation)},'' in {\em Optical Modeling and System Alignment},  M.~A.
	Kahan, J.~Sasi{\'a}n, and R.~N. Youngworth, Eds.,  {\bf 11103}, 111030N,
	International Society for Optics and Photonics, SPIE  (2019).
	
	\bibitem{Wyrowski_NanoOpt_JOSAA2020}
	R.~Shi, N.~Janunts, C.~Hellmann, {\em et~al.}, ``Vectorial physical-optics
	modeling of fourier microscopy systems in nanooptics,'' {\em Journal of the
		Optical Society of America. A, Optics, image science, and vision} {\bf
		37}(7), 1193–1205  (2020).
	
	\bibitem{Wyrowski_PolarizFieldTr_SPIE2019}
	S.~Zhang, C.~Hellmann, and F.~Wyrowski, ``{Polarization effects modeling with
		field tracing (Conference Presentation)},'' in {\em Polarization Science and
		Remote Sensing IX},  J.~M. Craven, J.~A. Shaw, and F.~Snik, Eds.,  {\bf
		11132}, 111320G, International Society for Optics and Photonics, SPIE
	(2019).
	
	\bibitem{Wyrowski_Seamless_SPIE24}
	F.~Wyrowski, ``{Seamless transition to geometrical optics concepts in a fully
		physical optics framework},'' in {\em Computational Optics 2024},  D.~G.
	Smith and A.~Erdmann, Eds.,  {\bf PC13023}, PC130230C, International Society
	for Optics and Photonics, SPIE  (2024).
	
	\bibitem{Wyrowski_LPIA_AplOpt00}
	A.~v. Pfeil, F.~Wyrowski, A.~Drauschke, {\em et~al.}, ``Analysis of optical
	elements with the local plane-interface approximation,'' {\em Applied Optics}
	{\bf 39}(19), 3304–3313  (2000).
	
	\bibitem{Malbet_EFC95}
	F.~{Malbet}, J.~W. {Yu}, and M.~{Shao}, ``{High-Dynamic-Range Imaging Using a
		Deformable Mirror for Space Coronography},'' {\em Pub. Astron. Soc. Pacific}
	{\bf 107}, 386  (1995).
	
	\bibitem{Moon&Stirling}
	T.~K. {Moon} and W.~C. {Stirling}, {\em Mathematical Methods and Algorithms for
		Signal Processing}, Prentice Hall  (2000).
	
\end{thebibliography}

\end{document}